# AmoebaContact and GDFold: a new pipeline for rapid prediction of protein structures

Wenzhi Mao[1,2], Wenze Ding[1,2] and Haipeng Gong[1,2,*]

[1]MOE Key Laboratory of Bioinformatics, School of Life Sciences, Tsinghua University, Beijing 100084, China.

[2]Beijing Advanced Innovation Center for Structural Biology, Tsinghua University, Beijing 100084, China.

*To whom correspondence should be addressed:

　　Email: hgong@tsinghua.edu.cn (H. G.)

**Abstract:** Native contacts between residues could be predicted from the amino acid sequence of proteins, and the predicted contact information could assist the *de novo* protein structure prediction. Here, we present a novel pipeline of a residue contact predictor AmoebaContact and a contact-assisted folder GDFold for rapid protein structure prediction. Unlike mainstream contact predictors that utilize human-designed neural networks, AmoebaContact adopts a set of network architectures that are found as optimal for contact prediction through automatic searching and predicts the residue contacts at a series of cutoffs. Different from conventional contact-assisted folders that only use top-scored contact pairs, GDFold considers all residue pairs from the prediction results of AmoebaContact in a differentiable loss function and optimizes the atom coordinates using the gradient descent algorithm. Combination of AmoebaContact and GDFold allows quick reconstruction of the protein structure, with comparable model quality to the state-of-the-art protein structure prediction methods.

# Introduction

As one of the most challenging problems in structural bioinformatics, prediction of protein structure from the amino acid sequence is of great importance and in urgent demand, especially with the unprecedentedly accumulated sequence data nowadays. The native contacts between residue pairs contain sufficient information for reconstructing the protein structure, and even knowledge of partial contact information could significantly accelerate protein folding simulations by effectively constraining the conformational search. Residue contacts could also be predicted from the amino acid sequence through sequence co-evolution analysis, for instance, using low-rank estimation methods like inverse covariance matrix and pseudo-likelihood maximization [1-4], and the prediction results typically constitute a square matrix listing the estimated contacting probabilities for all residue pairs, which is called the contact map. The contact map could be treated as a gray-scale image, which could be handled by deep learning and computer vision algorithms like the convolutional neural network (CNN). Moreover, pairs of fragments with similar relative postures in the three-dimensional structure usually exhibit similar local contact patterns on the residue contact map. This translational invariance property guarantees that computer vision algorithms are intrinsically appropriate for residue contact prediction. Accordingly, many deep-learning-based algorithms have emerged to further improve the accuracy of residue contact prediction, exemplified by the famous RaptorX-Contact [5, 6] that utilizes deep residual network (ResNet) for prediction. As a crucial resort to facilitate the protein structure prediction, protein residue contact prediction has become a regular project that has attracted many groups to participate in the recent critical assessment of techniques for protein structure prediction (CASP) competitions [6-8].

Despite the exciting progress, deep learning algorithms still confront some limitations in the residue contact prediction. Firstly, contact prediction is not a purely local pattern recognition problem. Indeed, the residue contact map is only a high-dimensional projection of the three-dimensional protein structure, for which the intrinsic degree of freedom is $3L$-6 (when counting the residue-based vibrational movements), where $L$ is the protein length. Therefore, the apparently independent $L^2$ or $L(L-1)/2$ residue pairs should be highly constrained within the contact map. For instance, according to a *priori* biological knowledge, each residue can only be in contact with at most 6-8 residues due to steric restriction [9], which imposes a strong limit on the number of contacting residue pairs in each row/column of the contact map. Similar sparsity constraints have been utilized to improve the prediction of β-β contacts by shallow learning algorithms like bbcontacts [10] and RDb$_2$C [11], which more or less engage the ranking information of sorted raw prediction scores for further selection. However, most deep learning algorithms do not take full advantage of such a *priori* biological knowledge, because the biological knowledge always involves some complex logics like sorting that are not GPU-friendly in computation and thus may significantly slow down the training, and more importantly because some operations like the sorted ranking are not differentiable and thus hinder the back-propagation gradient estimation.

Secondly, most neural-network-based algorithms simply borrow mature neural network architectures from the computer vision field for residue contact prediction, such that the particularity of residue contact problem may be potentially ignored. Neural architecture search (NAS), the emerging subfield of deep learning that tries to find more suitable architectures for specific tasks, may provide solutions to this problem. NAS algorithms could be described from 3 dimensions: search space, search strategy and performance estimation strategy [12]. Although search space should include all candidate architectures in principle, most NAS algorithms adopt cell-based or module-based architectures under the cost consideration [13, 14]. Search strategy, as the key of NAS algorithms that defines how to explore the search space in detail, must provide a balance of exploring efficiency and exploring coverage. NAS algorithms also develop skills to estimate the performance of candidate architectures [14-19] in order to significantly reduce the training cost and maintain the performance. Nowadays, most NAS algorithms could be classified as Bayesian optimization [16, 20, 21], evolutionary methods [18, 22, 23], reinforcement learning (RL) [13, 14, 24, 25] or gradient-based methods [26]. Recently, evolutionary methods and RL have shown exciting progresses. Particularly, the evolution-based AmoebaNet [18] exhibits considerable performance with much simpler searching strategy.

Conventionally, predicted residue contact maps are used to fold the protein by two types of contact-assisted folding algorithms. The first class of methods like C-QUARK [27] take the residue pairs of very high confidence from the predicted contact maps and integrate them as pseudo energy terms in traditional protein structure prediction algorithms to guide conformational sampling. The second class of methods like CONFOLD [28, 29] and RaptorX-Contact [6] set up constraining distance matrices using a large number of top-scored predictions and then run molecular dynamics simulations to find compatible protein structures. Interestingly, the latest version of RaptorX-Contact [6] that folds the protein from the predicted distance distribution of residue pairs (instead of the conventional 8 Å definition in contact prediction) is reported to achieve even better performance than traditional *de novo* structure prediction algorithms. In the latest CASP13, the conventional contact-assisted folding methods, however, have been challenged by AlphaFold [30], which utilizes multiple deep neural networks to predict the constraints like residue contacts and backbone torsion angles and more impressively folds the protein structures with these constraints following a simple gradient-descent-based scheme. Unfortunately, technical details of AlphaFold have not been fully released yet.

In this study, we present a new pipeline for rapid protein structure prediction, which consists of a novel contact predictor AmoebaContact and a gradient-descent-based contact-assisted folder GDFold. In contact prediction, we first introduced 2 new normalization operations as approximates of ranking to impose more internal constraints in deep learning frameworks. Moreover, we modified the AmoebaNet NAS algorithm to automatically search customized neural network architectures for the task of residue contact prediction. We also generalized the model to different contact cutoffs to gather more comprehensive distance information of residue pairs. Given the prediction results of AmoebaContact, we proposed a fast gradient-descent-based contact-assisted folding algorithm GDFold. Different from conventional contact-assisted folders that only use top-scored predictions, GDFold considers the overall predicted contact map in the differentiable loss function, which allows the optimization of atom coordinates using the gradient descent algorithm. The new pipeline of AmoebaContact and GDFold could produce protein structure models of comparable quality to the state-of-the-art method RaptorX-Contact but at a much faster speed.

## Results and Discussion

We utilized a well-performed NAS algorithm, AmoebaNet, to search optimal network architectures for protein residue contact prediction, but made the following modifications: (1) we incorporated the row normalization (RN) and column normalization (CN) into the cell-based structure of AmoebaNet; (2) we added the ResNet-like shortcut connections between adjacent layers to avoid the learning saturation; (3) we adjusted the list of candidate operations; (4) we accelerated the training by allowing inheritance of model weights during evolution. In total, we selected 15 architectures out of all explored networks. These selected models were then augmented and integrated into ensembles to improve the prediction of residue contacts at the cutoff of 8 Å.

As shown in Fig. 1, the models optimized at the cutoff of 8 Å were fine-tuned to a series of contact cutoffs to produce multiple predictors. The multiple outputs of AmoebaContact provide more comprehensive distance information for residue pairs and thus allow more efficient and accurate structure modeling. We designed a gradient-decent-based folding algorithm GDFold, which automatically optimizes the atom coordinates by minimizing a comprehensive differentiable loss function that imposes constraints of both predicted residue contacts and polypeptide geometry.

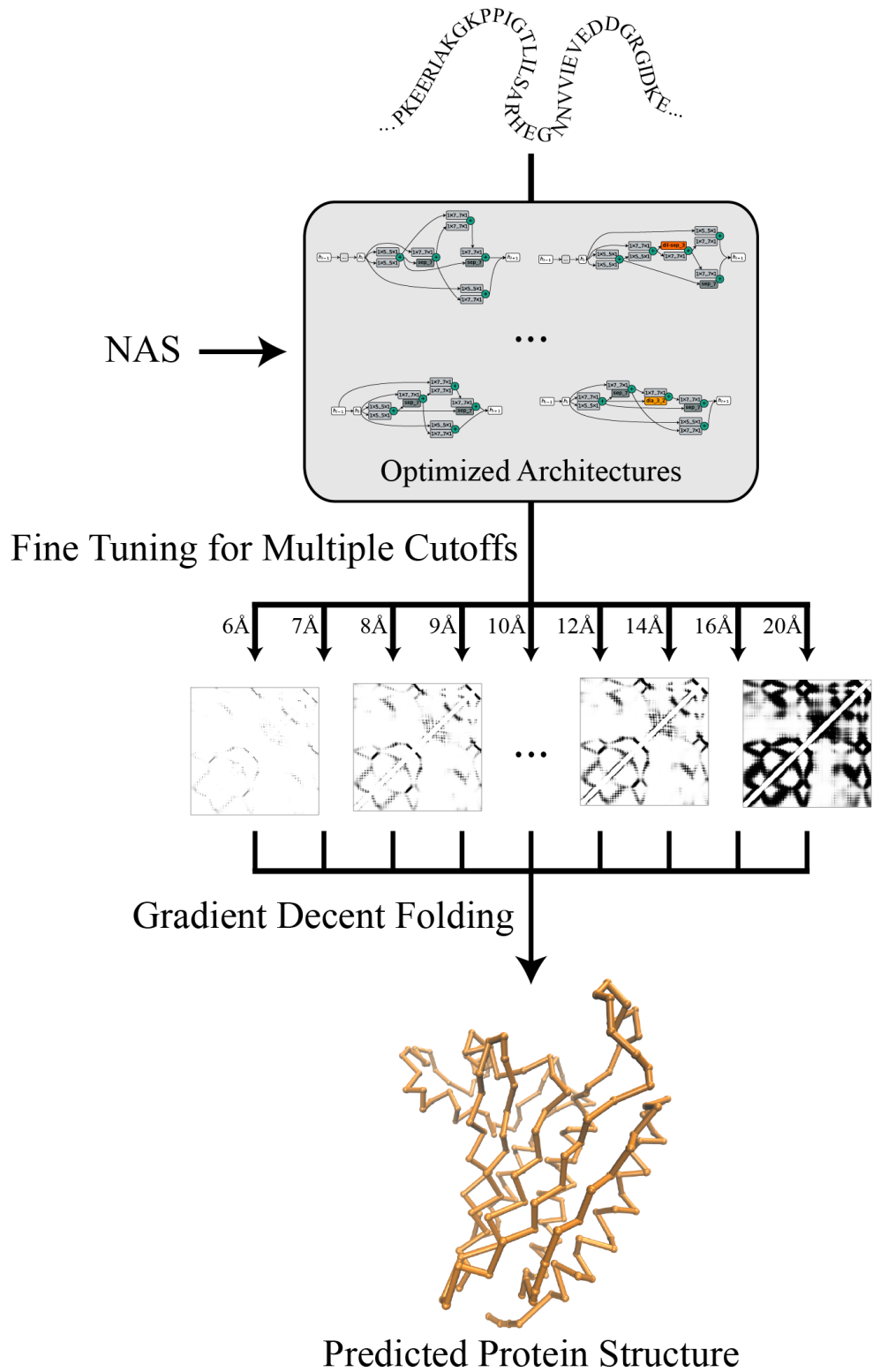

**Fig. 1. The general flow chart of the pipeline of AmoebaContact and GDFold.**

## Performance Improvement by RN and CN

We introduced 2 new operations, RN and CN, in the neural network as approximates of ranking to mimic the internal constraints of contacts around one residue. To validate the effectiveness of these operations in residue contact prediction, we conducted 3 sets of control experiments (Fig. S1). In experiment A, we constructed normal CNN and ResNet networks: each CNN layer contains a 3×3 convolution, followed by an instance normalization (IN) and a leaky ReLU (leaky rectified linear unit) activation [31], whereas each ResNet unit includes two sequential repeats of IN, leaky ReLU and 3×3 convolution as proposed by He et al. [32]. In experiment B, we replaced all IN with I/R/CN, a combination operation that performs parallel IN, RN and CN operations and then concatenates the normalization outputs. Considering that channels and thus trainable parameters are also triplicated by the I/R/CN operation, we applied three independent IN operations in parallel in experiment C as a control for fair evaluation. In all networks, 1×1 convolution layers are added at the beginning and end if necessary.

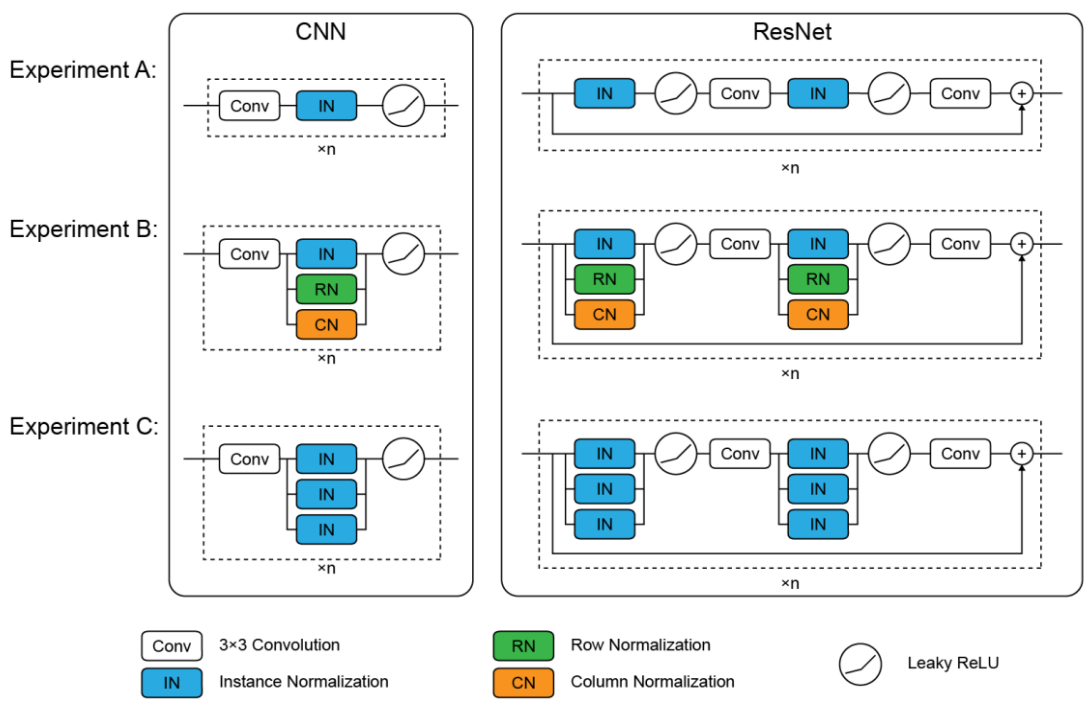

**Fig. S1. Network structure units for evaluating the contribution of RN and CN.**
In experiment A, we utilized the normal CNN and ResNet structures. In experiment B, we replaced IN by I/R/CN operation. In experiment C, we applied 3 independent IN operations in parallel as a control.

In all experiments, we fixed the number of channels to 30 and the contact cutoff to 8 Å. Equivalent CNN/ResNet networks were constructed for 10/5, 20/10 and 30/15 layers/blocks, respectively, were trained by 200 epochs on the training set, and were then evaluated by F1-score on the validation set. As shown in Fig. S2, Experiments A and C show very similar performance, whereas Experiment B significantly outperforms experiments A and C, which supports the positive contribution of RN and CN in residue contact prediction. Moreover, the superiority of I/R/CN in Experiment B is consistent in both neural architectures and for all network depths ever tested, and thus is unlikely to vanish in deeper networks. Therefore, the combination I/R/CN operation as designed here may further improve the residue contact prediction in other neural networks.

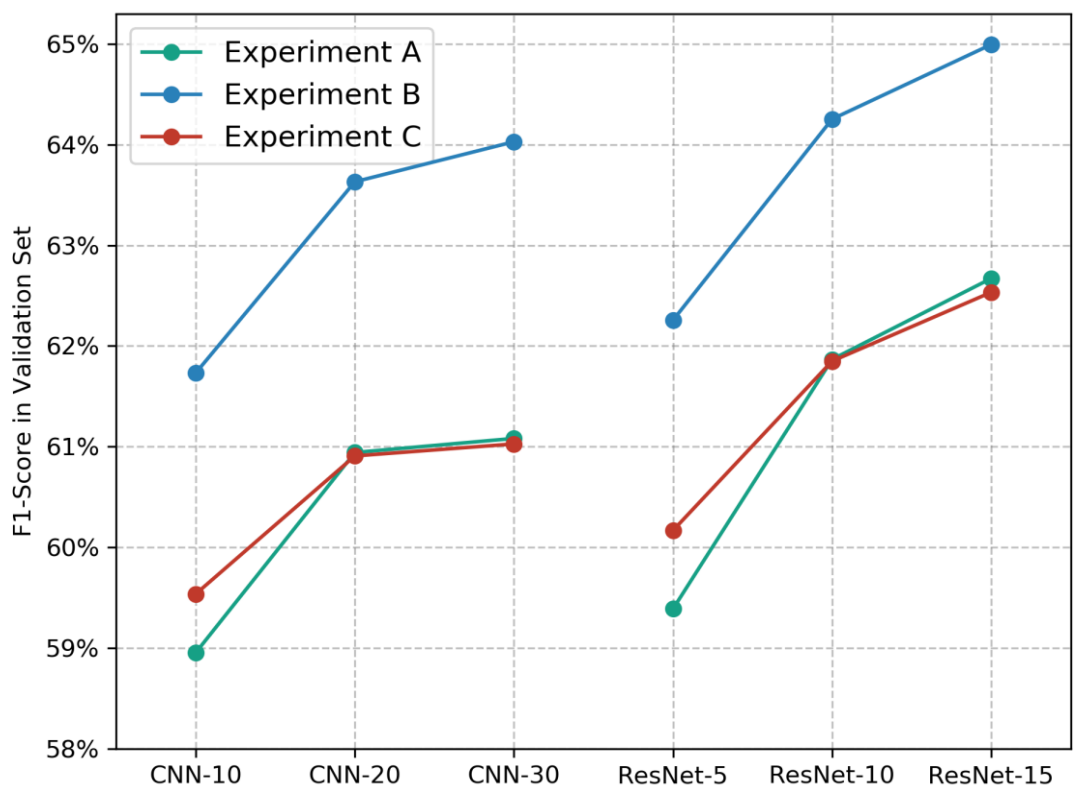

**Fig. S2. Performance evaluation for different normalization schemes.**
F1-scores of Experiments A, B and C are shown in green, blue and red, respectively. The evaluation was conducted for CNN of 10, 20 and 30 layers, and for equivalent ResNet of 5, 10, and 15 blocks.

**Neural Architecture Search**

As described previously, we modified the original AmoebaNet pipeline for residue contact prediction, including the introduction of I/R/CN operations. Using the modified algorithm, we explored a total of 500 network architectures, where the first 64 architectures were generated randomly while subsequent models were propagated from random architectures. Fig. 2A shows the F1-scores of the 500 architectures on the validation set. At the end of architecture search, the model performance converges to ~62%, with the best model reaching 62.16%.

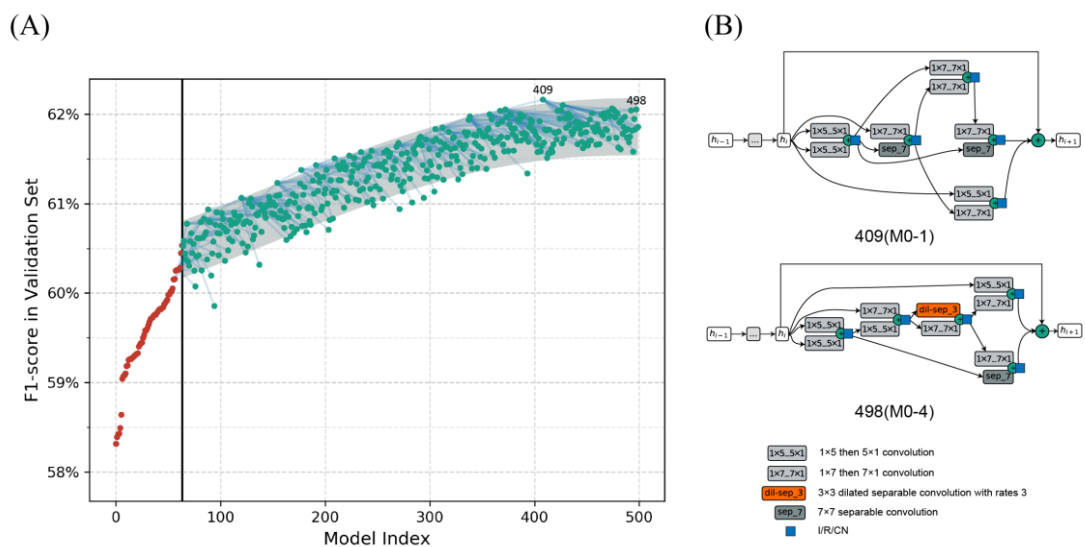

**Fig. 2. The model evolution in the AmoebaNet architecture searching process.**
**(A)** The horizontal axis represents the architecture index in the exploring process, and the vertical axis represents the F1-score on the validation set. The initial random architectures are marked as red dots, while the generated architectures are marked as green dots. The light blue line represents the inheritance relationship between architectures. Shaded region represents the general trend of performance evolution. **(B)** Examples of found architectures: model 409 (M0-1) and model 498 (M0-4).

Fig. S3 shows the change of operation composition during the evolution. As expected, proportions of all 17 operations are nearly identical in the first 64 models, but vary tremendously in evolution under the rule of survival of the fittest. 3 operations survive to the end, including 1×7→7×1 (1×7 convolution followed by 7×1 convolution), 1×5→5×1 (1×5 convolution followed by 5×1 convolution) and 7×7 separable convolution. The survived operations are all parameter-efficient, capable of enlarging the receptive field with minimal parameters. By using these operations, evolved models could avoid overfitting and improve the computational efficiency.

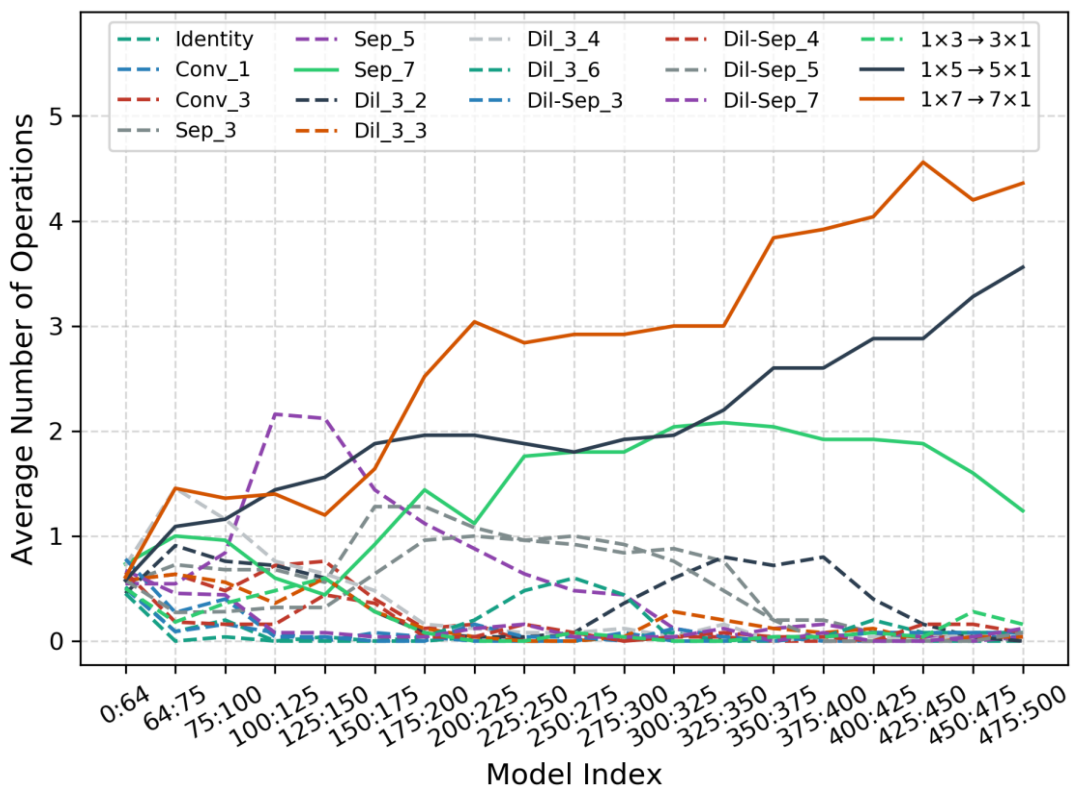

**Fig. S3. Change in the proportions of operations during the evolution process.**
The evolution process is divided into 19 periods, where the initial 64 random models are denoted as period 1 and each subsequent period contains 25 models. Operations with the final average number of > 1 are indicated by solid lines, while the others are indicated by dashed lines.

Among the 500 explored architectures, we selected 15 models for further optimization. Since the architectures with close inheritance relationship can hardly provide complementary information from each other, we must balance model performance and inheritance relationship. We first selected the best 5 models by F1-score as the M0 models (M0-1 to M0-5). After removing all models with a distance of 1 from all selected models on the genealogical tree, we collected the next 5 models by F1-score as M1 models (M1-1 to M1-5). To guarantee the model heterogeneity, we eliminated all models within a distance of 2 from selected models on the genealogical tree, and chose the last 5 models by F1-score as M2 models (M2-1 to M2-5). The performance of M0, M1 and M2 models ranges from 61.95% to 62.16% (Table S1 and Fig. S4). The detailed architectures of selected models are illustrated in Fig. 2B and Fig. S5.

**Table S1. General information of the M0, M1 and M2 models.**

| Selected Model | Model Index | F1-score on Validation Set |
| --- | --- | --- |
| M0-1 | 409 | 62.16% |
| M0-2 | 428 | 62.10% |
| M0-3 | 475 | 62.07% |
| M0-4 | 498 | 62.06% |
| M0-5 | 373 | 62.05% |
| M1-1 | 493 | 62.04% |
| M1-2 | 368 | 62.04% |
| M1-3 | 471 | 62.02% |
| M1-4 | 480 | 62.01% |
| M1-5 | 387 | 62.00% |
| M2-1 | 378 | 62.00% |
| M2-2 | 390 | 61.97% |
| M2-3 | 385 | 61.95% |
| M2-4 | 360 | 61.95% |

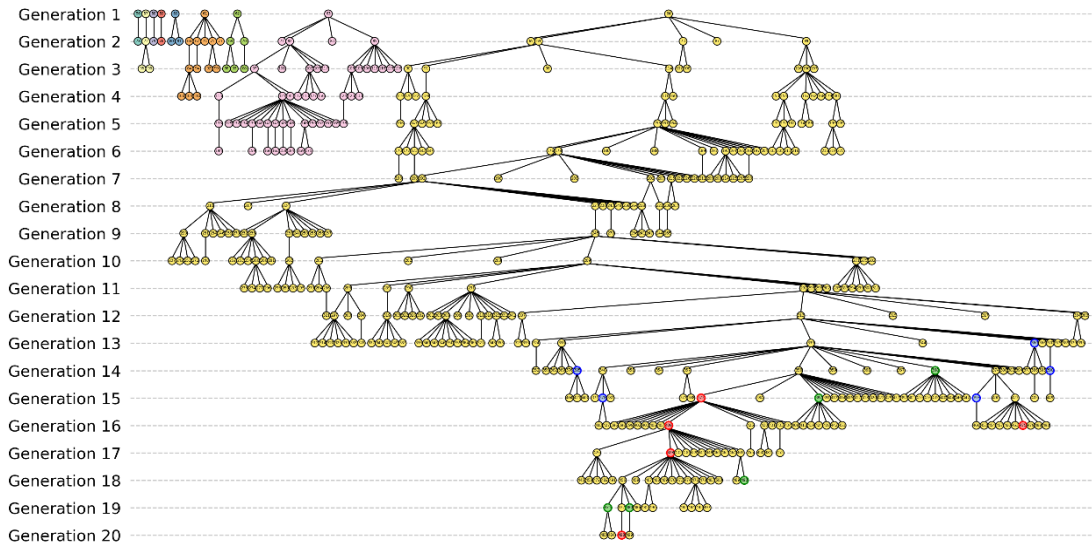

**Fig. S4. The inheritance relationship of architectures explored by AmoebaNet.**
The genealogical tree of all 500 explored architectures is shown here. Each node represents a network architecture, and the inheritance relationship between architectures is identified by an edge. Generation 1 refers to the initial random architectures, and following generations are defined by the inheritance. Architectures rooted from the same ancestor are labeled by the same colors. Initial random architectures with no descendants are omitted in this figure. The M0, M1 and M2 models are marked by red, green, and blue nodes, respectively.

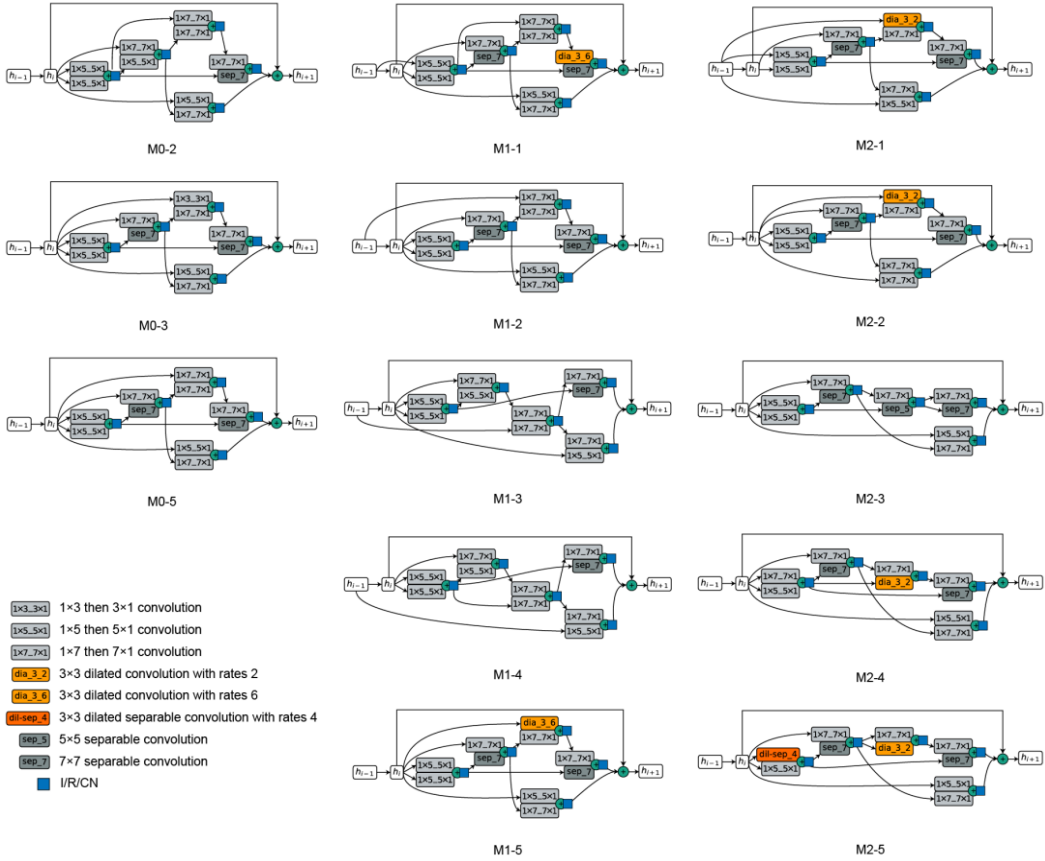

**Fig. S5. Cell architectures of selected models.**

Here we show the basic cell structures of selected models other than M0-1 and M0-4 (shown in Fig. 2B).

**Equivalent Model Analysis**

To validate the role of architecture searching, we tested the performance of equivalent ResNet models (with I/R/CN implemented) that have the same amounts of parameters to our selected models. For a ResNet with $N$ blocks and $F$ channels, the amounts of convolution parameters (Conv Para) and normalization parameters (Norm Para) are:

$$\begin{cases} \text{Conv Para} = 132F + 1 + 2NF(27F + 1) \\ \text{Norm Para} = 12NF \end{cases}$$

The hyper-parameters (web depth $N$ and channel number $F$) of equivalent ResNet models could thus be determined by equating the amounts of convolution and normalization parameters. As shown in Fig. S6, M0, M1 and M2 models are approximately equivalent to a ResNet with 7 blocks and 13-14 channels. Thereby, we selected 8 sets of ResNet hyper-parameters to conduct the independent training and evaluation. The equivalent depths and channel numbers were rounded upward to avoid underestimation of equivalent models.

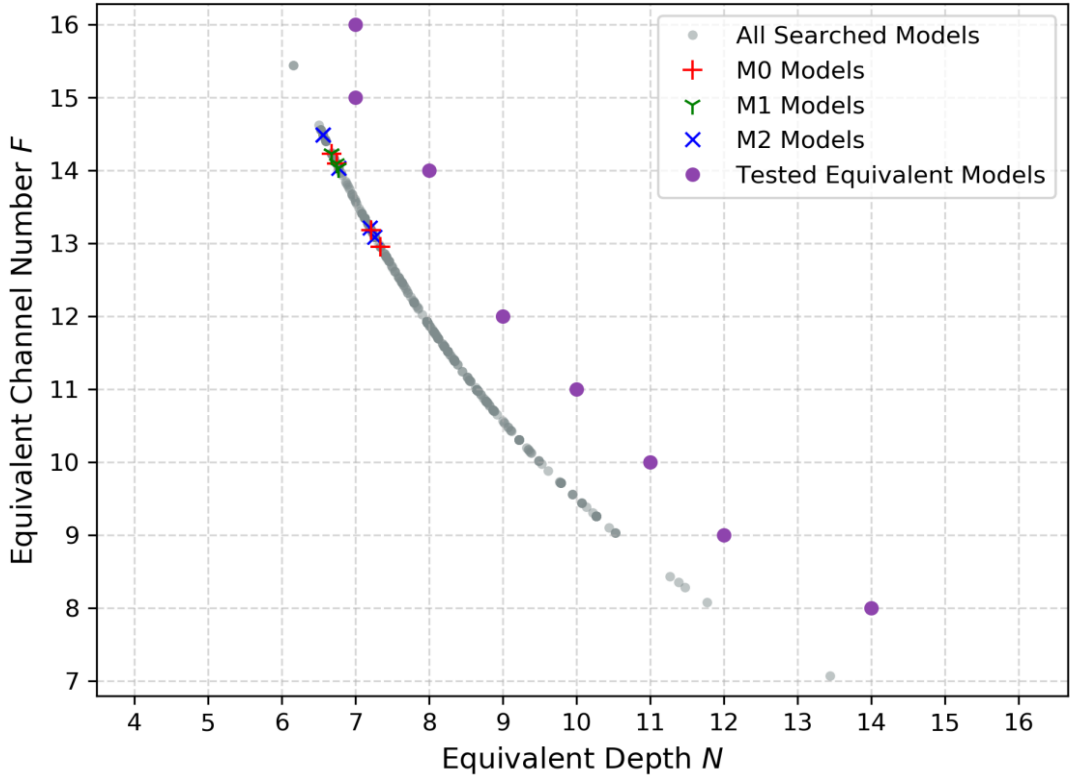

**Fig. S6. The hyper-parameters of equivalent ResNet models and the explored architectures.**
The selected M0, M1 and M2 models are labeled as red +, green Y and blue × symbols, respectively, while the other explored models are labeled as gray dots. The tested equivalent ResNet models are labeled as purple dots.

As listed in Table S2, the F1-scores of equivalent models range from 59.98% to 60.41%, even lower than the best random architecture generated in architecture search (60.53%). When comparing to the best optimized architecture (M0-1, 62.16%), the difference is further enlarged. These results support the significant performance gain brought by NAS.

**Table S2. Comparison of equivalent models and NAS explored models.**

| Equivalent Depth $N$ | Equivalent Channel Number $F$ | F1-score on Validation Set |
|---|---|---|
| 7 | 15 | 60.41% |
| 7 | 16 | 60.48% |
| 8 | 14 | 60.57% |
| 9 | 12 | 60.21% |
| 10 | 11 | 60.33% |
| 11 | 10 | 60.10% |
| 12 | 9 | 60.13% |
| 14 | 8 | 59.98% |
| Best Random Architecture | | 60.53% |

|              |         |
|--------------|---------|
| M0-1         | **62.16%** |

Best model is highlighted in bold.

## Model Augmentation

During the architecture search, the models were only trained on a small scale (3 blocks and 10 channels) under the consideration of computation efficiency, and they should be augmented for better performance. We augmented the models from two aspects: model depth $N$ and the number of channels $F$. Based on the 3-10 ($N$=3/$F$=10) models obtained in the architecture search, we carried out a series of augmentation experiments, by increasing $N$ to 4, 5 and 6 and $F$ to 30, 45 and 60, respectively. Despite the failure in training the M2-5 model for one hyper-parameter set (5-30 denoting $N$=5/$F$=30) due to memory overflow, the performance of all selected models generally improves upon model augmentation on both model depth and channel number (Fig. 3). Notably, model augmentation is still limited by GPU memory (e.g., 11GB used here), and more performance improvement is expected when further augmentation is allowed by more advanced hardware.

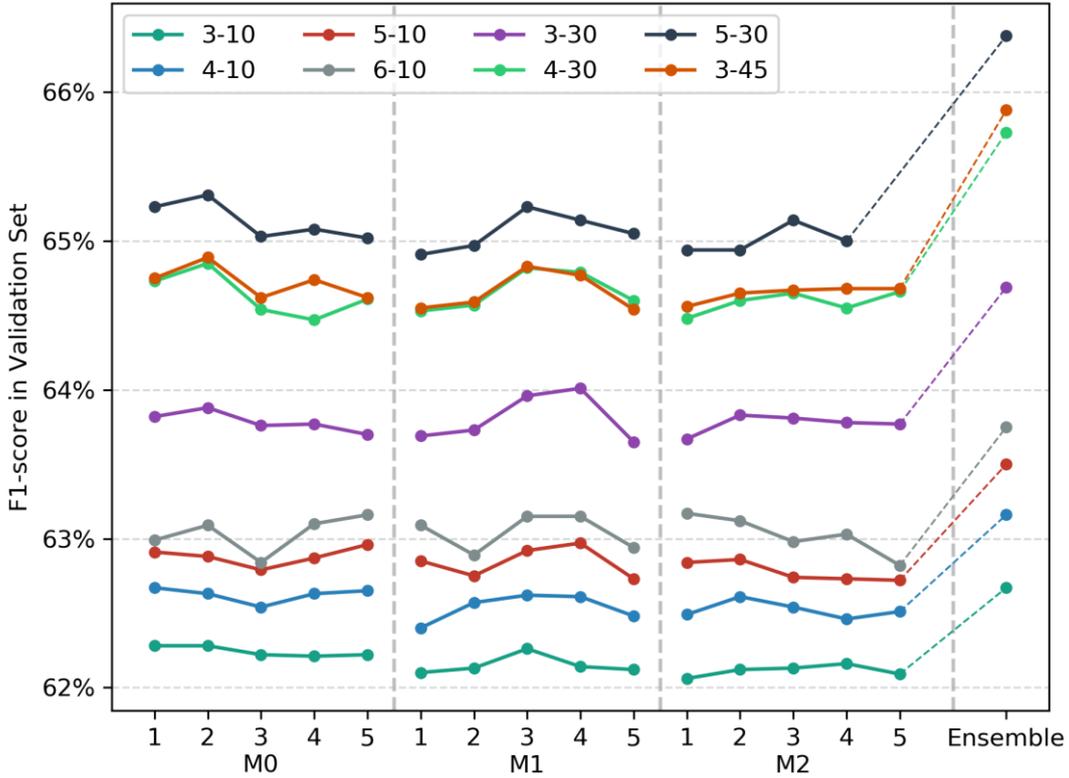

**Fig. 3. The performance of augmented models.**
The horizontal axis represents all 15 selected models (M0-1 to M0-5, M1-1 to M1-5 and M2-1 to M2-5) and the ensemble models. The vertical axis represents the F1-score on the validation set. Each curve in the figure represents a hyper-parameter combination. In the legend, the hyper-parameter combination is expressed as "network depth ($N$) - channel number ($F$)".

For each set of hyper-parameters, we sorted all M0, M1 and M2 models by F1-score on the validation set, and then selected the best $k$ models to build an ensemble model. We considered all possible $k$ values and chose the optimal model number $k_{best}$ on the validation set. Constructing an ensemble model can elicit a further performance improvement of ~1 percentage point in comparison to single models. The performance of the 5-30 ($N=5/F=30$) model is the best among all ensemble models, with the F1-score reaching 66.38%. We finally selected the 14 models of this hyper-parameter set (M2-5 model failed in training) for further parameter fine-tuning. In comparison to the un-augmented models, the model augmentation and ensemble averaging jointly enhance the F1-score from 62.16% to 66.38%.

**Fine-tuning for Multiple Contact Cutoffs**

All previous models were trained for predicting the residue contacts at a single cutoff ($C_\beta$ atom distance < 8 Å by conventional definition of contacts). However, outputs of these single-cutoff models are unlikely to provide comprehensive information of the distances between residue pairs, which is determinant in the three-dimensional structure modeling. Therefore, we fine-tuned the 14 models derived previously by architecture searching and model augmentation at a series of contact cutoffs (6, 7, 8, 9, 10, 12, 14, 16 and 20 Å).

Table S3. F1-scores on the validation set for models fine-tuned at various contact cutoffs.

| Cutoff | 6 Å | 7 Å | 8 Å | 9 Å | 10 Å | 12 Å | 14 Å | 16 Å | 20 Å |
|---|---|---|---|---|---|---|---|---|---|
| M0-1 | 56.38% | 61.44% | 65.40% | 65.60% | 66.50% | 73.26% | 76.25% | 78.95% | 83.43% |
| M0-2 | 56.48% | 61.42% | 65.44% | 65.72% | 66.63% | 73.33% | 76.31% | 78.96% | 83.41% |
| M0-3 | 56.17% | 61.16% | 65.23% | 65.39% | 66.20% | 72.97% | 75.96% | 78.67% | 83.18% |
| M0-4 | 56.26% | 61.24% | 65.25% | 65.42% | 66.33% | 73.12% | 76.10% | 78.90% | 83.29% |
| M0-5 | 55.96% | 61.23% | 65.18% | 65.38% | 66.32% | 73.14% | 76.19% | 78.97% | 83.28% |
| M1-1 | 56.09% | 61.18% | 65.03% | 65.30% | 66.25% | 73.05% | 76.12% | 78.89% | 83.32% |
| M1-2 | 56.20% | 61.16% | 65.12% | 65.38% | 66.31% | 73.01% | 76.04% | 78.78% | 83.20% |
| M1-3 | 56.58% | 61.59% | 65.38% | 65.64% | 66.70% | 73.36% | 76.32% | 78.94% | 83.29% |
| M1-4 | 56.45% | 61.38% | 65.25% | 65.55% | 66.54% | 73.21% | 76.24% | 78.90% | 83.34% |
| M1-5 | 56.20% | 61.35% | 65.20% | 65.51% | 66.43% | 73.19% | 76.20% | 78.96% | 83.41% |
| M2-1 | 56.22% | 61.16% | 65.08% | 65.20% | 66.24% | 72.98% | 76.01% | 78.68% | 83.21% |
| M2-2 | 56.18% | 61.17% | 65.05% | 65.28% | 66.22% | 73.07% | 76.04% | 78.76% | 83.13% |
| M2-3 | 56.18% | 61.31% | 65.27% | 65.49% | 66.39% | 73.20% | 76.22% | 78.90% | 83.35% |
| M2-4 | 56.30% | 61.24% | 65.18% | 65.41% | 66.43% | 73.10% | 76.23% | 78.94% | 83.39% |

As shown in Table S3, the performance of all 14 models improves with the rise of contact cutoffs, particularly for cutoffs over 10 Å. This phenomenon is mainly attributed to the drastic change in the positive-to-negative (Pos/Neg) ratio in all samples. At low cutoffs, majority of residue pairs are negative non-contact samples. With the growth of contact cutoff, the Pos/Neg ratio rises from ~1:120 to ~1:1.6 (Table S4), which greatly simplifies the model training for complicate neural networks.

**Table S4. The Pos/Neg ratio at different contact cutoffs.**

| Cutoff | Pos/Neg Ratio in Training Set | Pos/Neg Ratio in Validation Set |
|---|---|---|
| 6 Å | 1:121.28 | 1:126.25 |
| 7 Å | 1:68.59 | 1:71.58 |
| 8 Å | 1:43.15 | 1:45.04 |
| 9 Å | 1:28.26 | 1:29.47 |
| 10 Å | 1:18.38 | 1:19.20 |
| 12 Å | 1:8.82 | 1:9.24 |
| 14 Å | 1:5.24 | 1:5.51 |
| 16 Å | 1:3.28 | 1:3.46 |
| 20 Å | 1:1.56 | 1:1.66 |

We tested the powers of models optimized at various cutoffs for predicting contacts at other cutoffs (Table S5). As expected, each individual model outperforms the others at the cutoff they are trained for, whereas the performance deteriorates significantly when the tested cutoff is far away from the trained cutoff. Thus, models trained at different contact cutoffs are likely to provide different structural information.

**Table S5. Similarity of models trained for different contact cutoffs.**

| | | The Cutoff Tested on | | | | | | | | |
|---|---|---|---|---|---|---|---|---|---|---|
| | | 6 Å | 7 Å | 8 Å | 9 Å | 10 Å | 12 Å | 14 Å | 16 Å | 20 Å |
| The Cutoff Trained on | 6 Å | **57.90%** | 60.11% | 58.74% | 56.37% | 53.38% | 44.18% | 35.64% | 36.63% | 54.71% |
| | 7 Å | 55.26% | **62.80%** | 64.74% | 61.52% | 57.47% | 51.93% | 45.43% | 37.99% | 54.71% |
| | 8 Å | 49.14% | 60.26% | **66.65%** | 64.86% | 61.13% | 56.62% | 53.97% | 47.77% | 54.71% |
| | 9 Å | 45.37% | 57.52% | 65.45% | **67.01%** | 65.60% | 62.87% | 62.10% | 58.88% | 54.71% |
| | 10 Å | 42.79% | 54.09% | 62.01% | 65.45% | **68.03%** | 68.20% | 67.47% | 65.94% | 58.81% |
| | 12 Å | 31.41% | 43.14% | 50.80% | 54.91% | 61.47% | **74.51%** | 74.90% | 73.03% | 70.44% |
| | 14 Å | 21.32% | 32.32% | 42.50% | 49.73% | 56.42% | 71.81% | **77.48%** | 77.78% | 76.40% |
| | 16 Å | 14.83% | 23.71% | 33.22% | 42.49% | 52.00% | 67.53% | 75.08% | **80.09%** | 81.04% |
| | 20 Å | 8.03% | 13.45% | 20.00% | 27.81% | 37.72% | 57.87% | 68.37% | 76.29% | **84.41%** |

This table shows the F1-scores of the models trained with various contact cutoffs, when tested for different contact cutoffs on the validation set. Higher performance is colored as red and lower is colored as blue. Winner for each tested cutoff (i.e. column) is highlighted in bold.

Training multiple independent models at different cutoffs is computationally expensive. As a control, we built two all-in-one models that predict the contacts of different cutoffs with one uniform network: non-weighted and weighted all-in-one models. Within the loss functions, predictions from all contact cutoffs are treated equally in the former but are properly weighted to a roughly equal level of contribution in the latter (Table S6). Unfortunately, both weighted and non-weighted models underperform the independently trained models (Fig. S7). Although the all-in-one models could provide a 9:1 model compression, after considering both computational cost and performance, we finally chose independently trained models to build our AmoebaContact predictor.

**Table S6. The weights used in the weighted all-in-one model.**

| Cutoff | Cross Entropy | Weight |
|---|---|---|
| 6 Å | 0.018475 | 2.903040 |
| 7 Å | 0.028075 | 1.910370 |
| 8 Å | 0.039656 | 1.352477 |
| 9 Å | 0.056763 | 0.944873 |
| 10 Å | 0.079594 | 0.673845 |
| 12 Å | 0.122658 | 0.437266 |
| 14 Å | 0.166352 | 0.322414 |
| 16 Å | 0.209874 | 0.255555 |
| 20 Å | 0.267958 | 0.200159 |

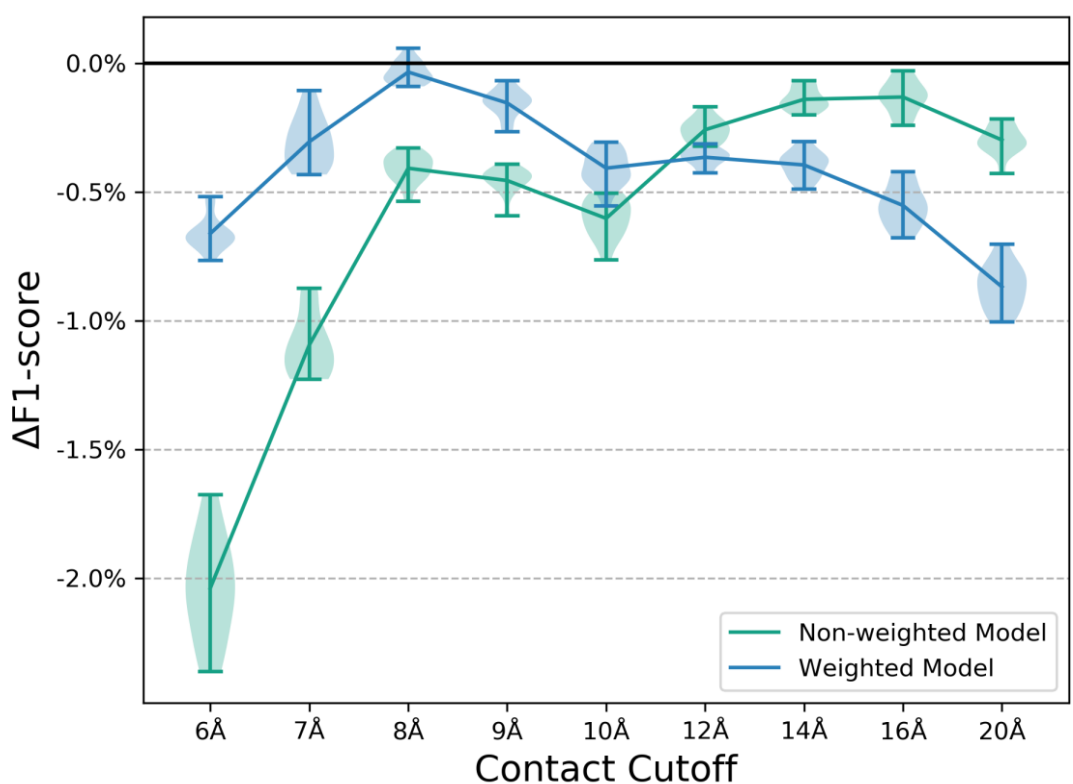

**Fig. S7. The performance of all-in-one models relative to individually trained models.**
The vertical axis represents the difference in F1-score (ΔF1-score) between the all-in-one model (green for non-weighted and blue for weighted) and the individually trained models. Negative values suggest that the performance declines in the all-in-one model. Solid lines describe comparison with the average of 14 individually trained models, while the detailed pairwise comparisons are shown in the violin-shaped shades.

The final AmoebaContact predictor consists of 9 ensemble models, each composed of 8-14 single models with different neural network architectures. As shown in Table 1, AmoebaContact exhibits balanced performance between the validation set and the PSICOV150 test set, but shows lower F1-scores in the three CASP test sets. This is not unexpected, because our contact predictor was trained on CATH domains [33] that represent the general prediction targets in practice, whereas the CASP targets are usually more difficult, with significantly less homologous sequences in the multiple sequence alignment (MSA) than proteins in the training, validation and PSICOV150 test sets (Fig. S8).

**Table 1. Performance of AmoebaContact.**

| Cutoff | Number of Models in Ensemble | Validation Set | PSICOV150 | CASP11 | CASP12 | CASP13 |
|---|---|---|---|---|---|---|
| 6 Å | 9 | 57.97% | 59.68% | 48.72% | 47.13% | 45.96% |
| 7 Å | 8 | 62.83% | 65.86% | 53.32% | 50.95% | 50.99% |
| 8 Å | 14 | 66.65% | 70.20% | 57.81% | 54.90% | 55.12% |
| 9 Å | 13 | 67.03% | 69.98% | 58.50% | 55.27% | 56.29% |
| 10 Å | 11 | 68.05% | 70.83% | 58.89% | 56.35% | 56.93% |
| 12 Å | 11 | 74.55% | 77.23% | 65.83% | 64.58% | 64.61% |
| 14 Å | 12 | 77.50% | 80.02% | 69.34% | 68.10% | 67.82% |
| 16 Å | 11 | 80.12% | 82.50% | 72.70% | 71.75% | 70.68% |
| 20 Å | 10 | 84.44% | 86.80% | 78.42% | 77.72% | 74.94% |

This table represents the F1-scores on the validation set and 4 test sets.

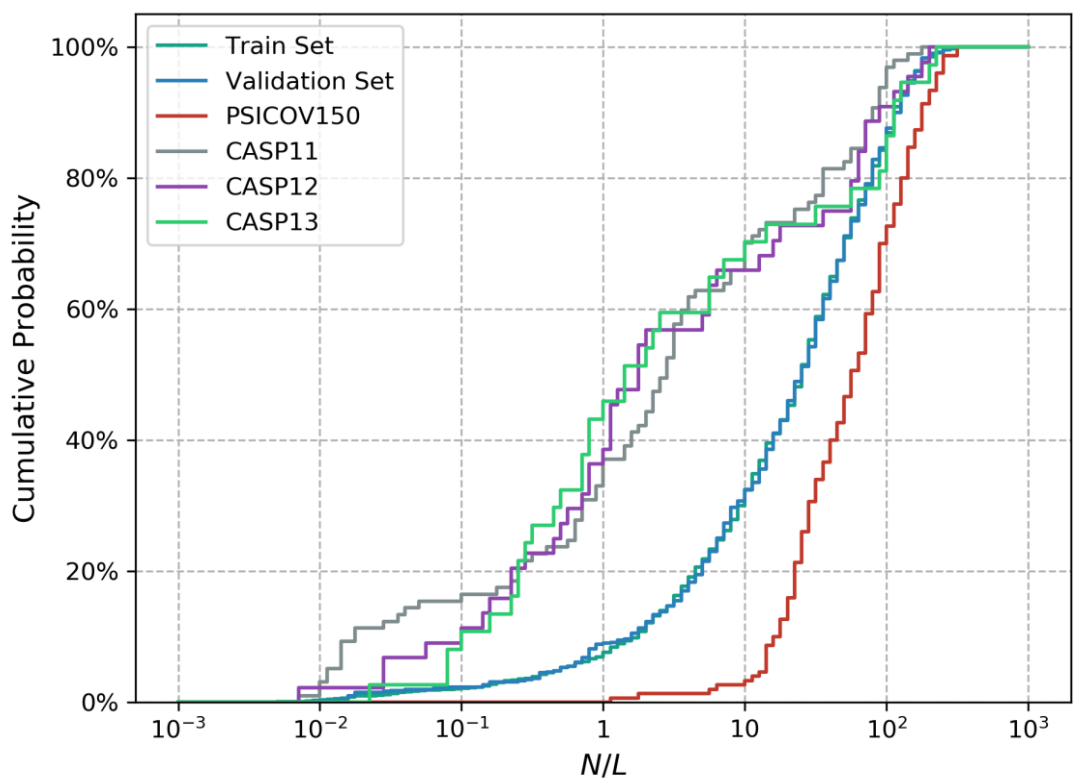

**Fig. S8. The cumulative probability distributions of protein targets in respect to the alignment depth of the MSA for the training, validation and test sets.**
*N* is the number of sequences in the MSA and *L* is the protein length. *N/L* is used to quantify the level of alignment depth.

**Structure Modeling by GDFold**

We first analyzed the relationship between the prediction score of AmoebaContact and the reliability of the score on the validation set. Specifically, we divided the range of prediction scores (0-1) into 20 equally sized bins, assigned all predictions into bins, and then calculated the average score as well as the proportion of positive samples for each bin. Interestingly, the proportion of the positive samples, a.k.a. the reliability of prediction, is almost identical to the prediction score for all AmoebaContact models (Fig. S9), indicating that the prediction score of AmoebaContact could be roughly regarded as an unbiased estimation of the probability of a residue pair being in contact. Hence, high-scored and low-scored predictions of AmoebaContact suggest the corresponding residue pairs are likely to be in contact and out of contact with high confidence, respectively. That is, both top-scored and non-top-scored predictions by our contact predictor may provide useful information for structure modeling.

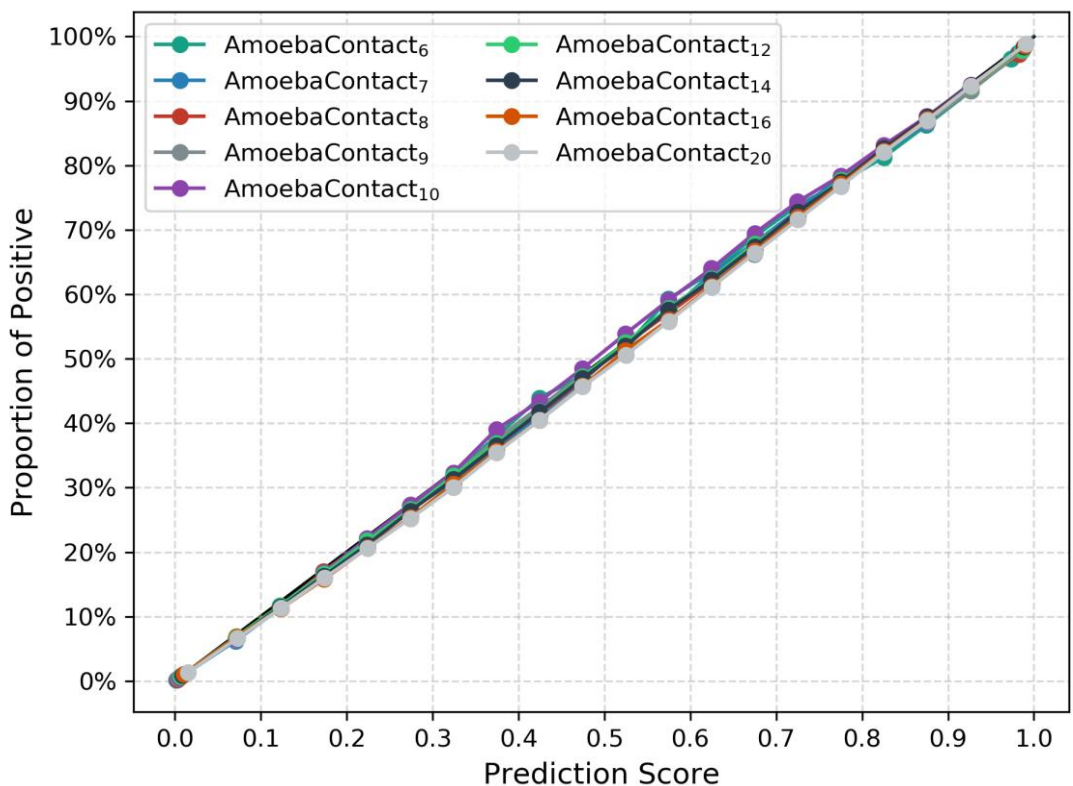

**Fig. S9. Relationship between the prediction scores and the proportions of positive samples for AmoebaContact models trained at various cutoffs.**
The range of prediction scores (0-1) is divided into 20 intervals with an identical width of 0.05. Each dot reflects the mean score and proportion of positive samples in one interval.

To fully utilize the information in the predicted contact map, we designed a gradient-decent-based folding algorithm GDFold to model the protein structure. Unlike other contact-assisted folding algorithms that mainly utilize top-scored residue pairs, GDFold adopts a differentiable scoring function to evaluate the consistency between atom coordinates and contact predictions. Since the scoring function is differentiable, gradient decent algorithm could be utilized to continuously update the atom coordinates so as to improve the structure-contact matching. Besides the contact-related terms, the scoring function also incorporates certain terms to impose the basic geometric constraints of polypeptides, like the distance between adjacent $C_\alpha$ atoms, excluded volume, handedness of helices, etc. Using GDFold, we could quickly obtain the protein structure that best fits the AmoebaContact prediction.

We evaluated the performance of GDFold against CONFOLD and RaptorX-Contact on the test sets. Structural modeling by RaptorX-Contact was performed through the latest web server (Mar. 2019). As for CONFOLD, we used various levels of top-scored predictions (from $0.5L$ to $3L$, where $L$ is the protein length) from AmoebaContact$_8$ (the ensemble model for the cutoff of 8 Å) as the distance restraints for structure modeling. In addition, we also fed predictions over the suggested cutoff of AmoebaContact$_8$ (optimized with F1-score on the validation set) to CONFOLD.

CONFOLD, RaptorX-Contact and GDFold all provided 5 structures as the final results, and the best RMSDs and TM-scores of the 5 models were averaged among targets in each test set. As shown in Table 2, GDFold outperforms CONFOLD in all test sets. Thus, predicted contact information from multiple cutoffs indeed facilitates structure modeling better than the traditional single-cutoff predictions. Generally, GDFold has a comparable performance to RaptorX-Contact: GDFold exhibits better performance in the PSICOV150 set but shows slightly weaker performance in the CASP sets.

**Table 2. Comparison of GDFold, CONFOLD and RaptorX-Contact.**

|  |  | PSICOV150 | CASP11 | CASP12 | CASP13 |
|---|---|---|---|---|---|
| AmoebaContact + CONFOLD | $0.5L$ | 12.55 Å/0.43 | 19.85 Å/0.33 | 17.91 Å/0.35 | 16.44 Å/0.34 |
|  | $1.0L$ | 7.19 Å/0.60 | 13.06 Å/0.43 | 11.37 Å/0.42 | 12.64 Å/0.41 |
|  | $1.5L$ | 5.33 Å/0.68 | 10.60 Å/0.48 | 10.08 Å/0.46 | 10.88 Å/0.46 |
|  | $2.0L$ | 4.50 Å/0.71 | 9.42 Å/0.51 | 9.34 Å/0.48 | 10.26 Å/0.48 |
|  | $2.5L$ | 4.34 Å/0.72 | 9.03 Å/0.51 | 9.03 Å/0.48 | 10.15 Å/0.47 |
|  | $3.0L$ | 4.23 Å/0.72 | 8.59 Å/0.51 | 9.13 Å/0.48 | 10.11 Å/0.47 |
|  | Cutoff | 4.68 Å/0.72 | 11.97 Å/0.50 | 12.77 Å/0.48 | 11.79 Å/0.47 |
| GDFold |  | **3.84 Å/0.74** | 7.76 Å/0.53 | 8.22 Å/0.51 | 9.39 Å/0.47 |
| RaptorX-Contact |  | 4.55 Å/0.72 | **7.58 Å/0.57** | **8.19 Å/0.58** | **8.97 Å/0.52** |

Average RMSDs and TM-scores are listed before and after the slash, respectively. Winner in each category is highlighted in bold.

We then systematically compared AmoebaContact$_8$ and GDFold against the contact prediction and structural modeling by RaptorX-Contact for individual proteins in all test sets (Fig. 4 and Fig. S10). In the PSICOV150 set, AmoebaContact$_8$ and GDFold both significantly outperform RaptorX-Contact. In the more difficult CASP test sets, AmoebaContact$_8$ is remarkably outperformed by RaptorX-Contact (Fig. 4A, Table S7 and Fig. S11). However, the inferiority of AmoebaContact is successfully rescued by GDFold in structure modeling. Structure models produced by GDFold and RaptorX-Contact show no obvious difference in RMSD (Fig. 4B) or TM-score (Fig. S10), and Deming regression analysis [34] denies the presence of significant difference in all CASP sets (Table S8).

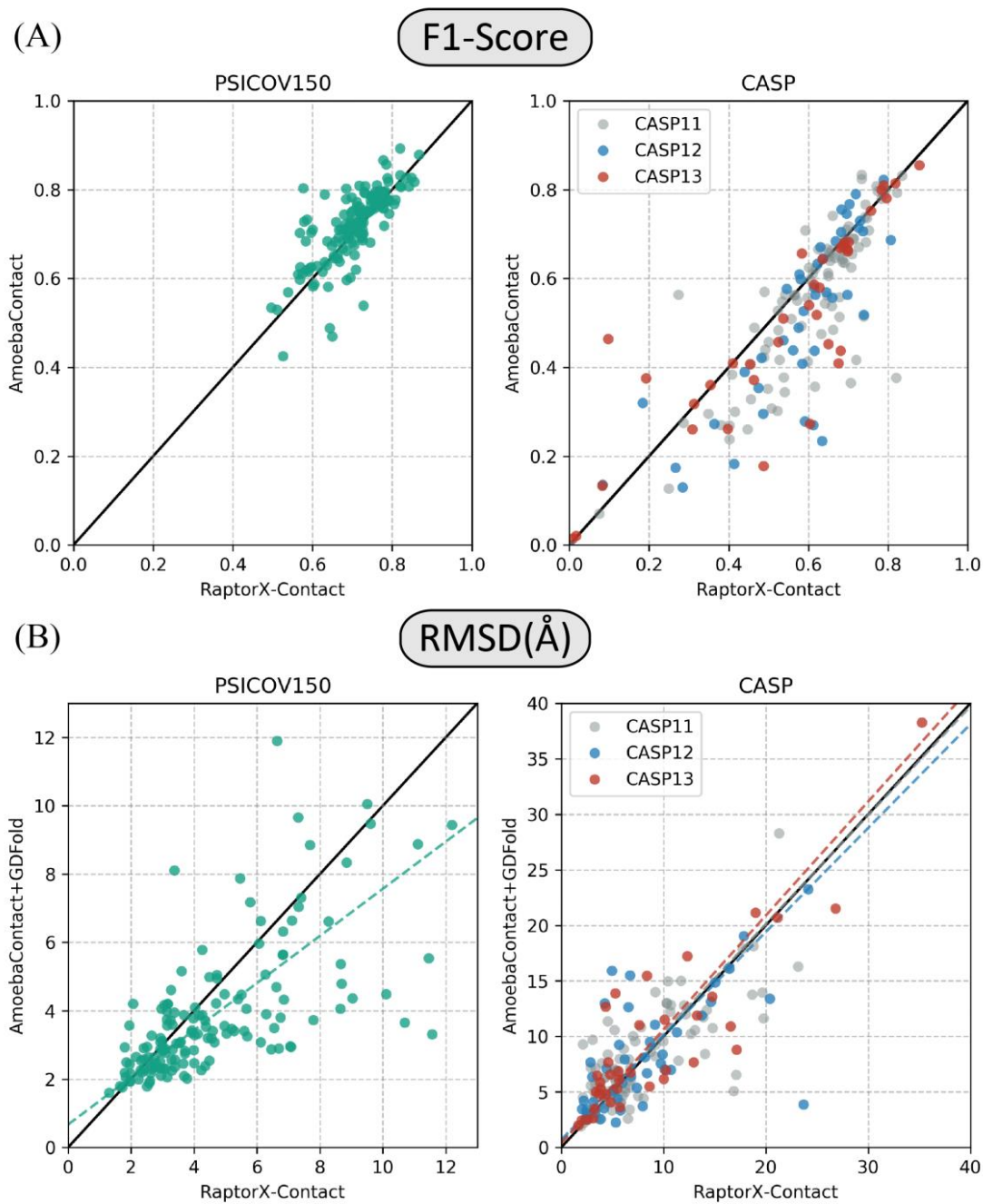

**Fig. 4. Protein-wise comparison of AmoebaContact and GDFold vs. the contact prediction and structure modeling by RaptorX-Contact in the test sets.**
**(A)** Comparison of contact prediction between RaptorX-Contact and AmoebaContact, where each dot describes the F1-scores of one individual protein target as evaluated by the two methods. **(B)** Comparison of model quality between RaptorX-Contact and AmoebaContact/GDFold pipeline, where each dot describes the best RMSD for one individual proteins target as evaluated by the two methods. The diagonal line is labeled as solid black, while the results of the Deming regression test are labeled by dashed lines.

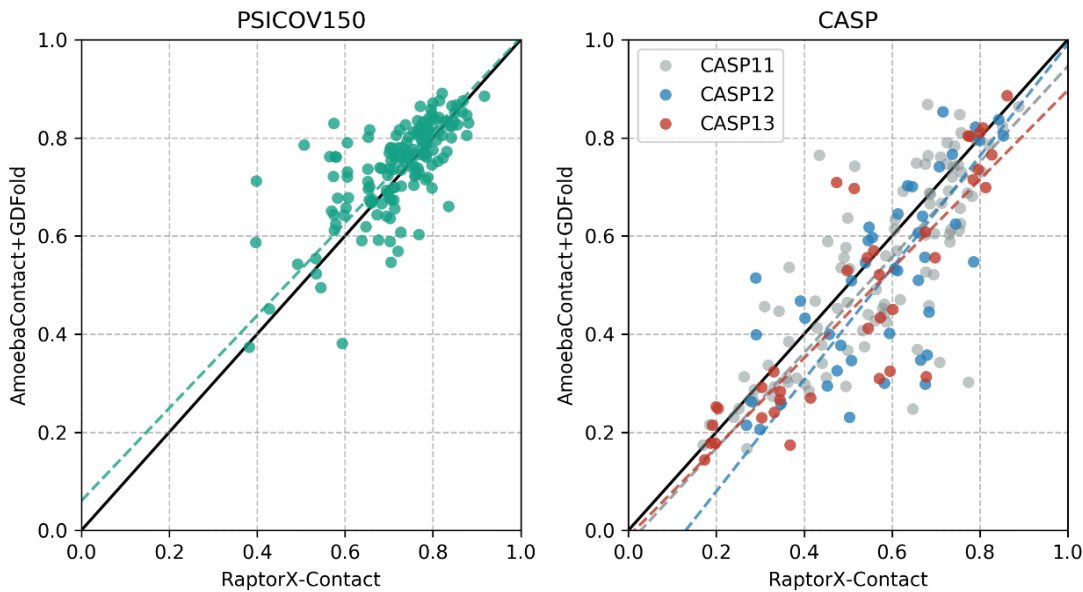

**Fig. S10. TM-score comparison of GDFold and RaptorX-Contact.**
The black line is the diagonal and the dashed lines are the results of Deming regression.

**Table S7. Precision of top-scored contact predictions by AmoebaContact and by RaptorX-Contact.**

|  |  | PSICOV150 | | CASP11 | | CASP12 | | CASP13 | |
|---|---|---|---|---|---|---|---|---|---|
|  |  | RC* | AC* | RC* | AC* | RC* | AC* | RC* | AC* |
| Short | $L$/10 | 90.3% | **94.5%** | 77.4% | **81.7%** | 74.6% | **76.4%** | 69.0% | **73.5%** |
|  | $L$/5 | 80.7% | **85.3%** | 70.2% | **71.8%** | **67.1%** | 66.6% | 61.1% | **62.0%** |
|  | $L$/2 | 52.5% | **55.2%** | **47.7%** | 47.6% | 43.8% | **44.3%** | 41.9% | **42.8%** |
|  | $L$ | 30.2% | **31.2%** | **29.1%** | 29.0% | 25.8% | **26.2%** | 26.9% | **27.1%** |
| Medium | $L$/10 | 90.3% | **94.6%** | **79.6%** | 79.0% | **75.8%** | 72.0% | **76.0%** | 67.6% |
|  | $L$/5 | 83.6% | **87.1%** | **73.5%** | 72.8% | **68.1%** | 64.7% | **66.9%** | 59.1% |
|  | $L$/2 | 62.3% | **64.0%** | **55.4%** | 54.1% | **48.7%** | 47.1% | **47.5%** | 46.3% |
|  | $L$ | 38.4% | **39.3%** | **36.6%** | 35.4% | **30.4%** | 29.3% | **32.2%** | 31.3% |
| Long | $L$/10 | 96.8% | **99.0%** | **79.4%** | 72.4% | **75.7%** | 66.7% | **66.4%** | 60.2% |
|  | $L$/5 | 94.5% | **97.0%** | **74.0%** | 68.1% | **74.1%** | 61.2% | **63.1%** | 56.0% |
|  | $L$/2 | 87.8% | **89.8%** | **66.1%** | 58.8% | **65.1%** | 51.3% | **56.0%** | 47.7% |
|  | $L$ | 74.9% | **75.4%** | **54.9%** | 47.6% | **52.4%** | 41.2% | **46.2%** | 39.6% |
| F1-score |  | 70.6% | **72.0%** | **58.9%** | 53.8% | **56.4%** | 49.8% | **53.8%** | 50.0% |

*RC and AC are the abbreviations of RaptorX-Contact and AmoebaContact, respectively. Winner in each category is highlighted in bold.

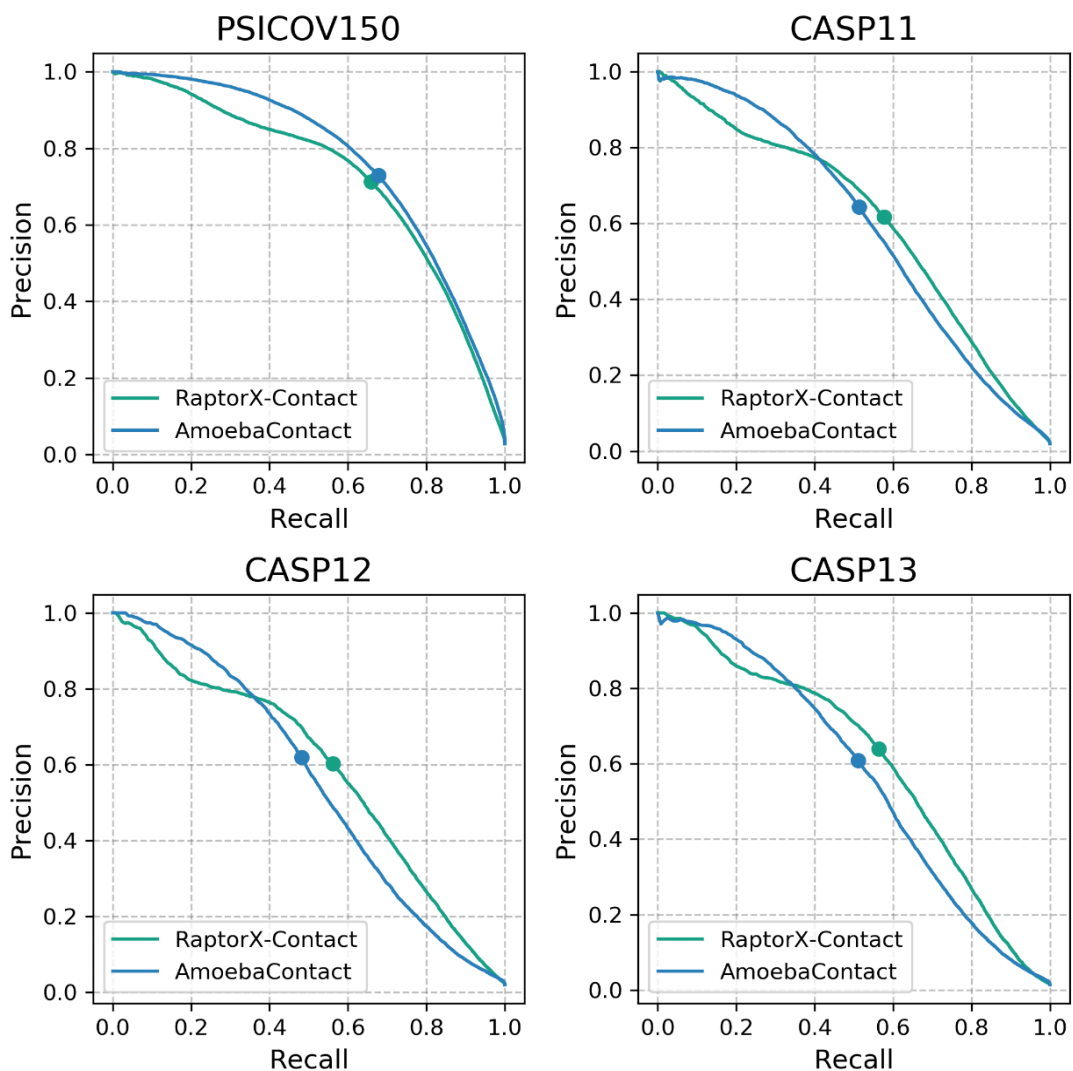

**Fig. S11. The precision-recall curves of AmoebaContact and RaptorX-Contact in the test sets.**
The comparison is shown for RaptorX-Contact (green) and AmoebaContact (blue) in the PSICOV150 and 3 CASP sets respectively. Best F1-scores are marked as dots on the curves.

**Table S8. Deming regression analysis for RMSD and TM-score between GDFold and RaptorX-Contact.**

| Indicator | Test Set | Parameter | Estimated values | Lower bound* | Upper bound* |
|---|---|---|---|---|---|
| RMSD | PSICOV150 | Slope | 0.690122 | 0.562856 | 0.817388 |
|  |  | Intercept | 0.669393 | 0.266901 | 1.071884 |
|  | CASP11 | Slope | 0.980360 | 0.810707 | 1.150014 |
|  |  | Intercept | 0.517009 | -0.312919 | 1.346937 |
|  | CASP12 | Slope | 0.935495 | 0.609864 | 1.261125 |
|  |  | Intercept | 0.704886 | -0.909200 | 2.318971 |
|  | CASP13 | Slope | 1.025996 | 0.800664 | 1.251328 |
|  |  | Intercept | 0.371154 | -0.340715 | 1.083022 |
| TM-score | PSICOV150 | Slope | 0.943891 | 0.663014 | 1.224767 |
|  |  | Intercept | 0.060175 | -0.147776 | 0.268126 |

|  |  |  |  |  |
|---|---|---|---|---|
| CASP11 | Slope | 0.973085 | 0.850038 | 1.096132 |
|  | Intercept | -0.027399 | -0.090869 | 0.036072 |
| CASP12 | Slope | 1.139447 | 0.858319 | 1.420576 |
|  | Intercept | -0.148707 | -0.320672 | 0.023258 |
| CASP13 | Slope | 0.909532 | 0.769255 | 1.049809 |
|  | Intercept | -0.012678 | -0.066867 | 0.041511 |

*with confidence of 95%.

We suspect that GDFold is capable of rescuing the inferiority of AmoebaContact predictions for CASP targets because the full predicted contact map is utilized for structure modeling. To test this idea, we kept the top $3L$ AmoebaContact predictions unchanged but shuffled the rest prediction scores. Such non-top-shuffled contact maps were then fed to GDFold for structure modeling. In both PSICOV150 and CASP sets, the non-top-shuffled contact prediction significantly deteriorates the quality of constructed structure models (Fig. S12 and Fig. S13), which reinforces the contribution of non-top-scored contact predictions in structure modeling.

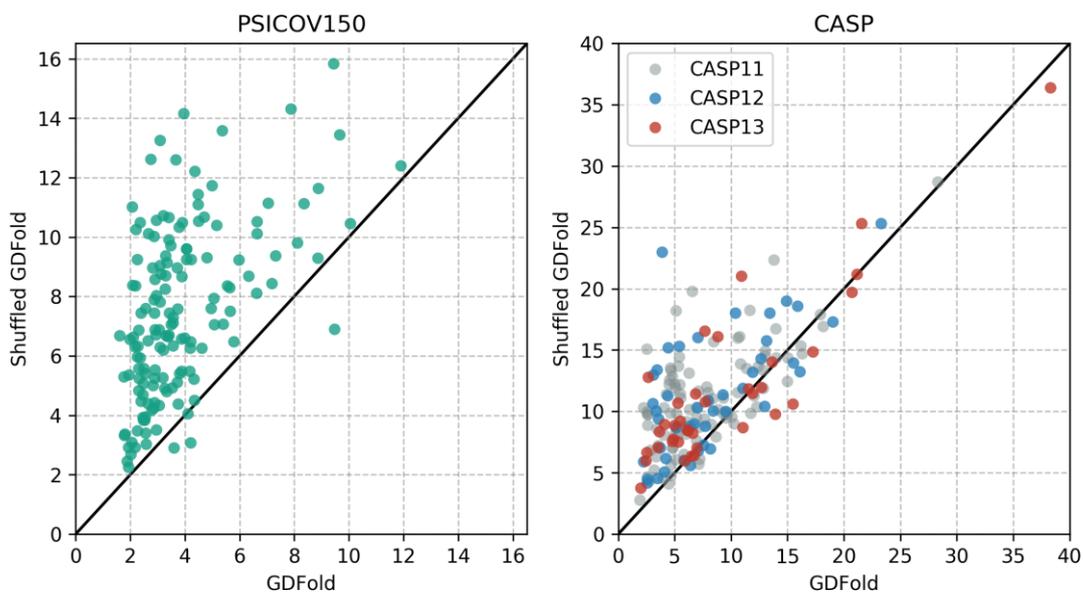

**Fig. S12. RMSD comparison of GDFold and non-top-shuffled GDFold.**

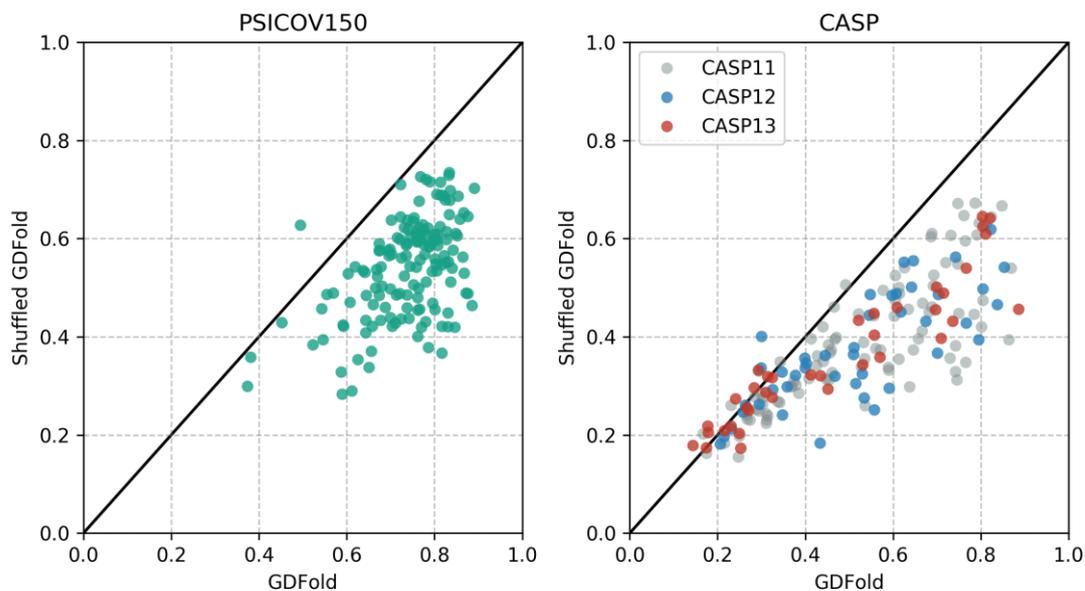

**Fig. S13. TM-score comparison of GDFold and non-top-shuffled GDFold.**

Fig. 5 shows the structure prediction of one protein case by our pipeline of AmoebaContact and GDFold, for which the overall topology of β-sheets and α-helices are correctly reconstructed. As a comparison, the RaptorX-Contact model has some misfolded region in β-sheets.

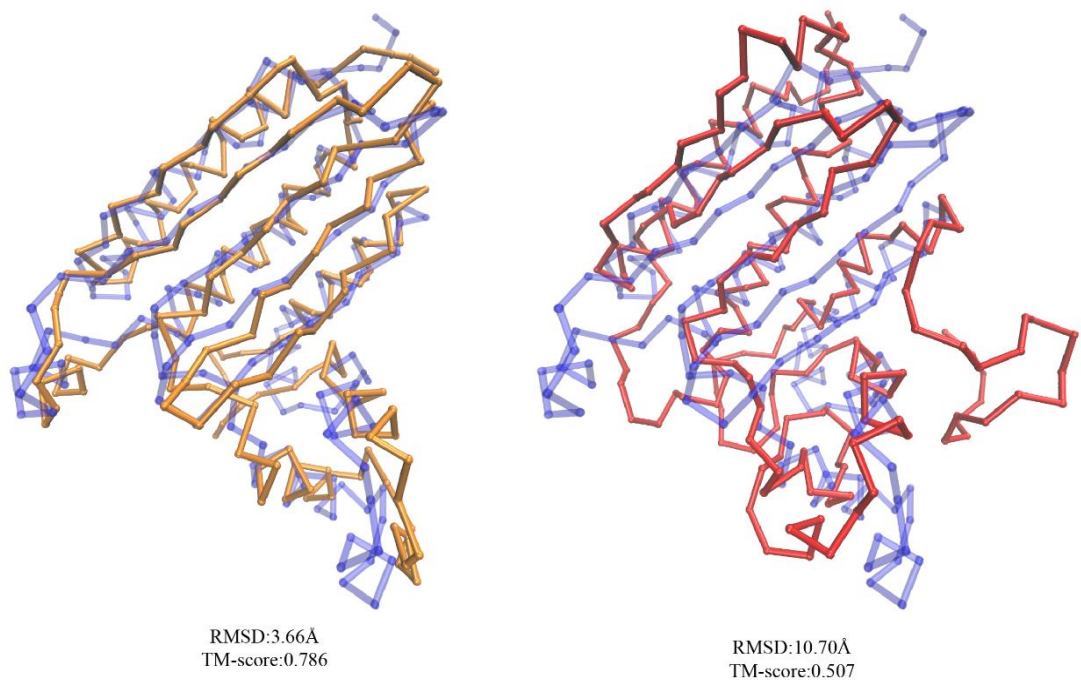

RMSD:3.66Å
TM-score:0.786

RMSD:10.70Å
TM-score:0.507

**Fig. 5. Case study.**
The models of $C_\alpha$ trace for the target 1I58A as predicted by AmoebaContact/GDFold (orange) and RaptorX-Contact (red), aligned with the native structure (blue). In this case, our pipeline reaches a smaller RMSD (3.66 Å *vs* 10.70 Å) and a higher TM-score (0.789 *vs* 0.507).

In conclusion, we employed the AmoebaNet to optimize the neural network architecture for protein residue contact prediction. The selected models were generalized to multiple contact cutoffs to obtain more comprehensive contact information. Our multi-cutoff contact predictor AmoebaContact could provide more useful information to infer protein structure. We then developed a gradient-decent-based folding algorithm GDFold to build the structure model that best fits the prediction results of AmoebaContact. Evaluation on the PSICOV150 and CASP sets suggest that combination of AmoebaContact and GDFold could generate protein structure models with comparable quality to the state-of-the-art method like RaptorX-Contact. Meanwhile, computational consumption of GDFold is significantly lower than CONFOLD (Fig. S14) and contact-assisted folders that rely on the CONFOLD protocol, e.g., RaptorX-Contact. Moreover, CONFOLD-dependent contact-assisted folders always need multiple runs to produce a large number of models using different levels of top-scored pairs, whereas GDFold avoids such selection by directly utilizing the overall contact map.

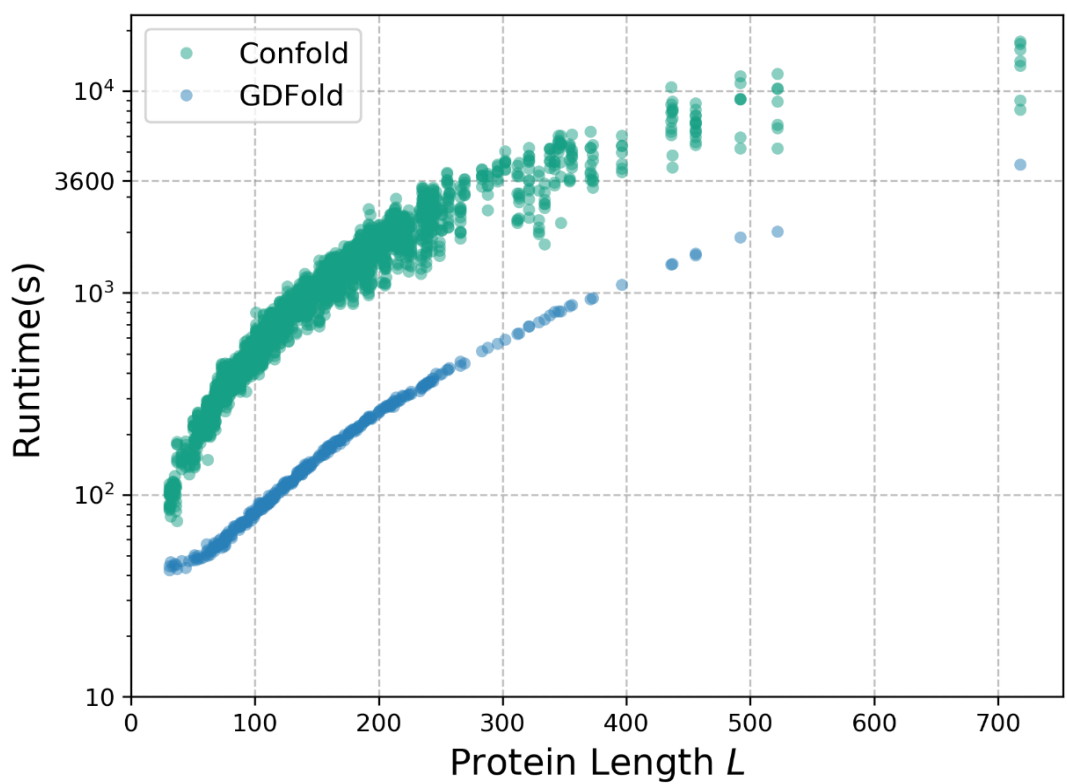

**Fig. S14. Runtime Comparison of CONFOLD and GDFold.**
The horizontal axis represents the length $L$ of target protein, and the vertical axis represents the runtime (in seconds) for structural prediction.

## Materials and Methods

### Dataset

We used several well-established datasets for testing, including PSICOV150 [2], CASP11 [35], CASP12 [8] and CASP13. The training and validation sets were derived from the CATH database of protein domain (version 4.1) [33]. Specifically, we kept all high-resolution structures (> 2.5 Å) in CATH as the start point. In order to remove the redundancy between the training/validation set and test sets, we eliminated all domains that belong to the same CATH fold groups as proteins in the test sets. The fragmented and very short (< 50 residues) or very long (> 500 residues) domains were also disregarded. At last, only domains in the CATH S35 set [36] (a subset of CATH with pairwise sequence identity < 35%) were retained to reduce the redundancy inside the training/validation set. Finally, there were 2994 domains left, and we randomly selected 20% of them (599 domains) as the validation set and kept the remaining as the training set (2395 domains) (Table S9).

Table S9. General information of the training set, validation set and test sets.

|  | Proteins | Residues | Residue Pairs | Contact Residue Pairs |
|---|---|---|---|---|
| Training Set | 2395 | 350569 | 30662892 | 694582 |
| Validation Set | 599 | 88807 | 8166168 | 177356 |
| PSICOV150 | 150 | 21838 | 1675968 | 47431 |
| CASP11 | 97 | 16342 | 1800056 | 34713 |
| CASP12 | 44 | 6853 | 658503 | 12484 |
| CASP13 | 37 | 5984 | 861883 | 12033 |

### Model Features

For all protein domains, we defined the contact of residue pairs by the distance between $C_\beta$ atoms ($C_\alpha$ for glycine): if the distance is less than a certain contact cutoff, the residue pair is regarded as contact, and vice versa.

We used multiple features to predict the residue contacts. The MSAs were first built by HHblits [37] against the UniProt20 database [38]. Then, the $L \times L$ residue contact maps were predicted by CCMpred [3] and MI [39], where $L$ is the protein length. We also adopted some predicted information by other programs to profile the property of each amino acid residue. The secondary structure probabilities were predicted by DeepCNF [40, 41] (3 classes and 8 classes, in total $L \times 11$). Additional residue properties (like $\phi$, $\psi$, $\theta$, $\tau$, etc.) were predicted by SPIDER3 [42] ($L \times 10$). The identities of amino acid residues were encoded by one-hot vectors ($L \times 20$). The frequency in evolution of each residue was derived from the corresponding columns in the MSA ($L \times 21$, where gap is counted as a type of residue). To allow the model to know the relative position between residues, we added the difference of residue indices as additional features. Occasionally, the MSA was highly segmented in multi-domain proteins. To conquer this problem, we also counted the probability of each residue pair to co-exist in one sequence in MSA. At last, we included the protein length ($L$) and the number of sequences in MSA ($N$) as features.

**Row Normalization and Column Normalization**

The proper normalization is very important in neural networks. In certain situations, the normalization is critical to solve the problem. For example, as a specific version of batch normalization [43], instance normalization [44] could bring significant quality improvements in the stylized image generation with only small network changes. Instance normalization (IN) was firstly introduced by Ulyanov and Vedaldi in 2016:

$$\text{IN}_{i,j,k} = \frac{Input_{i,j,k} - \mu_k}{\sigma_k} \gamma_k + \beta_k, \text{where} \begin{cases} \mu_k = \frac{1}{WH} \sum_{i=1}^{W} \sum_{j=1}^{H} Input_{i,j,k} \\ \sigma_k^2 = \frac{1}{WH} \sum_{i=1}^{W} \sum_{j=1}^{H} (Input_{i,j,k} - \mu_k)^2 \end{cases}$$

The key idea is that the data (like brightness or contrast) in an image may follow a specific distribution. As IN provides an approximation of the percentile, the global distribution fitting could be improved.

There are also some distribution limitations in protein contact prediction. Taking $\beta$-$\beta$ contact prediction as an example, each $\beta$ residue could only be in contact with at most 2 residues. This limitation has been well-utilized in some $\beta$-$\beta$ contact prediction algorithms like bbcontacts and RDb$_2$C: there are at most 2 positive labels in each row or column of contact matrices. There are also similar limitations for general residue contacts: each residue can only be in contact with at most 6-8 residues due to the steric restriction. In order to incorporate such distribution limitation of each row and column in the contact matrix, we introduced 2 new operations, row normalization (RN) and column normalization (CN), based on IN:

$$\text{RN}_{i,j,k} = \frac{Input_{i,j,k} - \mu_{j,k}}{\sigma_{j,k}} \gamma_k + \beta_k, \text{where} \begin{cases} \mu_{j,k} = \frac{1}{W} \sum_{i=1}^{W} Input_{i,j,k} \\ \sigma_{j,k}^2 = \frac{1}{W} \sum_{i=1}^{W} (Input_{i,j,k} - \mu_{j,k})^2 \end{cases}$$

$$\mathrm{CN}_{i,j,k} = \frac{Input_{i,j,k} - \mu_{i,k}}{\sigma_{i,k}} \gamma_k + \beta_k, \text{where} \begin{cases} \mu_{i,k} = \frac{1}{H} \sum_{j=1}^{H} Input_{i,j,k} \\ \sigma_{i,k}^2 = \frac{1}{H} \sum_{j=1}^{H} (Input_{i,j,k} - \mu_{i,k})^2 \end{cases}$$

As modifications of IN, RN and CN simply redefine the data range to estimate the parameter μ and σ². As illustrated in Fig. S15, RN and CN only utilize a row or a column to estimate the distribution parameter, unlike the whole channel in IN. Noticeably, the learnable parameters $\gamma$ and $\beta$ are not customized for each row or column. This is because the size of the contact map is not fixed.

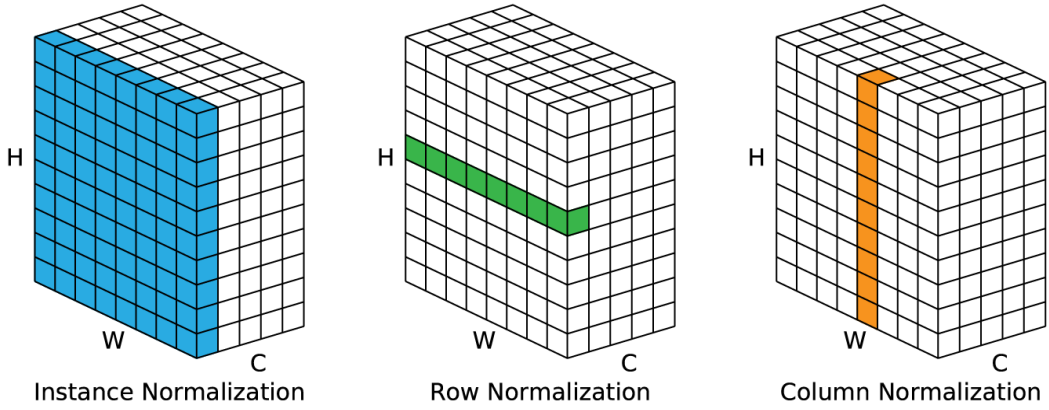

**Fig S15. Comparison of IN, RN and CN.**

The colored parts represent the corresponding region to estimate the mean and variance.

**Neural Architecture Search by AmoebaNet**

In this study, we adopted the AmoebaNet as NAS algorithm to optimize the network architecture. AmoebaNet is a cell-based NAS algorithm. There are two types of cells: normal cell and reduction cell. Each cell is a complicate sub-module and the whole network is built by stacking normal cells and reduction cells in a certain sequence. Normal cell and reduction cell could be regarded as a generalized convolution layer and a pooling layer, respectively. Normal cell will maintain the shape of input and reduction cell will reduce the output size by a factor of 2. In each cell, the outputs of previous two cells are taken as initial hidden states (*H0* and *H1*). More hidden states are then constructed through pairwise combinations. Each pairwise combination consists of applying an operation to an existing hidden state and applying another operation to another existing hidden state, where the operation is extracted from a predefined candidate operation list **O**. The sum of these two operations is regarded as a new hidden state and added to existing hidden state set. The pairwise combination will be repeated for *C* times. At last, all hidden states that have not been selected by previous pairwise combinations will be concatenated together, and a 1×1 convolution will be added as the output of this cell. In AmoebaNet, all normal cells have the same architecture, as are reduction cells. For all operations in normal and reduction cells, the number of channels (output filters) is fixed at $F$.

During the architecture evolution, a population of $P$ architectures will be maintained all the time. The population is randomly initialized at the beginning. Then, a subset of size $S$ will be sampled from the population. The model with the best validation performance will be selected as a parent. A new child architecture will be generated from the parent by a mutation of an operation type or a connection in the cell. The new child will be added into the population while the oldest architecture will be removed to maintain the population size unchanged. The evolution will be repeated until the performance converges or enough architectures have been generated. In order to optimize the architecture more efficiently, the number of channels $F$ and the number of normal cells $N$ will be reduced during the evolution. The selected model will be augmented (by increasing $F$ and $N$) after the evolution.

In this study, we have made several changes to the original AmoebaNet (Fig. S16). Firstly, our output is an $L \times L$ contact matrix instead of a single classification. So the reduction cell is not necessary in our case. We only need to optimize the normal cell, which simplifies our architecture search space. Secondly, candidate operation list **O** is also adjusted. We started with the SP-III operation list [18] introduced in AmoebaNet. We removed all pooling operations and added some dilated convolutions with dilation rate of 3 or 4 for the identification of the $\alpha$-helix cycle. Finally, we selected 17 operations as our candidate operation list **O**: identity mapping; 1×1, 3×3 convolutions (Conv_1, Conv_3); 3×3, 5×5, 7×7 separable convolutions (Sep_3, Sep_5 and Sep_7); 3×3 dilated convolutions with rates 2, 3, 4 and 6 (Dil_3_2, Dil_3_3, Dil_3_4 and Dil_3_6); 3×3 dilated separable convolutions with rates 3, 4, 5 and 7 (Dil_Sep_3, Dil_Sep_4, Dil_Sep_5 and Dil_Sep_7); 1×3 then 3×1 convolution (1×3→3×1); 1×5 then 5×1 convolution (1×5→5×1); 1×7 then 7×1 convolution (1×7→7×1). In trial experiments, we found that the model augmentation did not always bring performance improvement. To conquer this shortage, we added the ResNet-like skip connection in each cell to prevent performance deterioration. At last, we added the I/R/CN operation after each pairwise combination and the output of each cell to incorporate the row and column information. During training, we modified the original AmoebaNet strategy to allow each architecture to inherit the network weights from its parent. The weights for mutation paths or operations would be re-initialized.

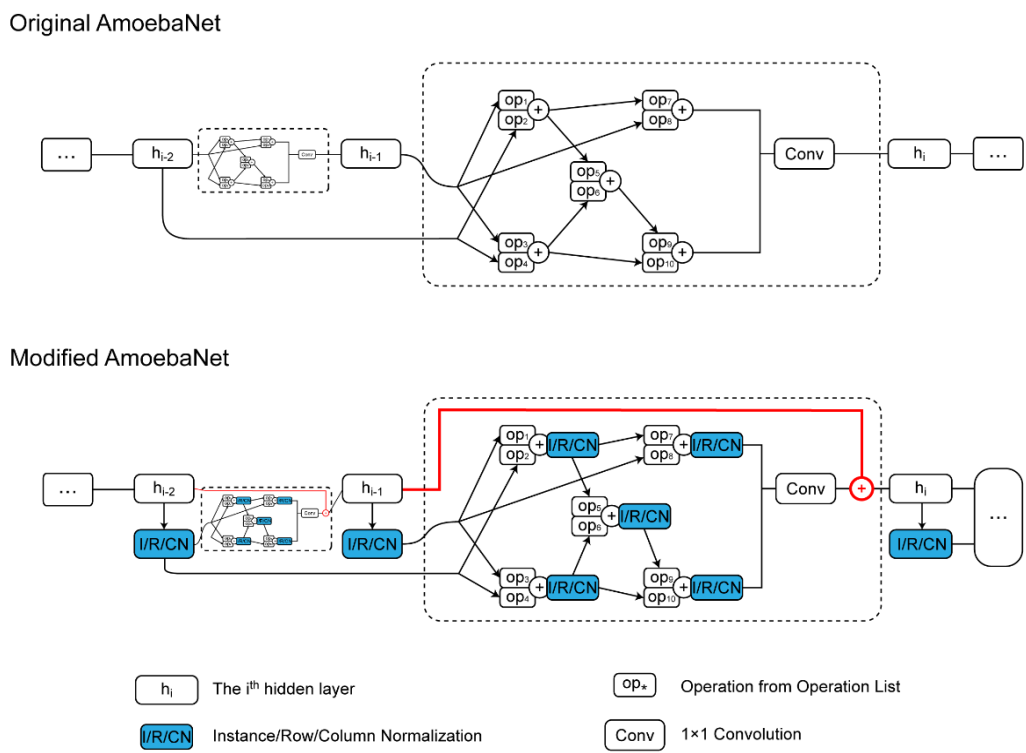

**Fig. S16. Network structure comparison of modified AmoebaNet and original AmoebaNet.**
We made two major modifications to the network structure of AmoebaNet. Firstly, we added the I/R/CN operation after each pairwise combination and cell output (the blue modules in the figure). Secondly, we also added the skip connection in each cell (red lines in the figure) to prevent the learning saturation.

In general, we searched for architectures with repetition of *N* normal cells. The number of pairwise combination *C* was 5 for each normal cell. During the architecture searching, the number of channels *F* and cell repeat number *N* were fixed at 10 and 3, respectively. Initial 64 architectures were generated randomly. The population size and re-sampling size were fixed at 64 and 16, respectively. Each architecture was trained for 100 epochs with a learning rate of $10^{-3}$. The contact cutoff was set as 8 Å during the searching phase. When a total of 500 architectures were explored, the performance had converged and no more significant improvement was observed.

**Model Augmentation and Fine-tuning**

After the architecture searching, the selected models (M0, M1 and M2) were augmented to larger and more accurate models. To accomplish this, we enlarged the cell repeat number *N* and channel number *F*. We tried a series of parameter combinations: *N* from 3 to 6 and *F* of 10, 30, 45 and 60. However, due to the hardware limitation, we could only enlarge the *F* to 45, 30, 30 and 10, when *N* is 3, 4, 5 and 6, respectively. For *N*=5/*F*=30, the training of M2-5 failed due to memory overflow.

After the determination of hyper-parameters (*N* and *F*), we fine-tuned the models for other contact cutoffs. The weights from 8 Å 5-30 ($N$=5/$F$=30) models were taken as initializations for other contact cutoffs training. Learning rate was optimized at $10^{-4}$, and the models were trained for 100-200 epochs for proper convergence (different for different cutoffs).

**Gradient Decent Folding**

The predicted contact maps of multiple cutoffs from AmoebaContact were utilized in GDFold. For each cutoff, the predicted contact map was transformed into a loss function as below:

$$L_{ij} = -P_{ij} \times \log(D_{ij}) - (1 - P_{ij}) \times \log(1 - D_{ij}),$$

where $P_{ij}$ is the prediction score of residue $i$ and residue $j$, and $D_{ij}$ represents whether residue $i$ and residue $j$ are in contact in the structure. To make $L_{ij}$ differentiable, $D_{ij}$ is softened by a sigmoid function.

$$D_{ij} = Sigmoid(d_{cutoff} - d_{ij}),$$

where $d_{ij}$ is the distance between residue $i$ and $j$ in angstrom and $d_{cutoff}$ refers to the cutoff value for the specific predicted contact map. To address the protein structure property better, we also added several other loss terms. We built a ResNet-based local contact predictor to predict the local ($|i\text{-}j|$<6) contact information (see Supplementary Materials for details), which was integrated in loss in a similar manner to non-local contacts. Since contact information could not prevent the handedness mismatch, we also developed a multi-layer perceptron model to predict whether a 4-residue fragment belongs to the $\alpha$-helix (see Supplementary Materials for details). Constraints were applied to local residues and the dihedral angle of adjacent $C_\alpha$ atoms to ensure proper handedness. Other constraints like distance of adjacent $C_\alpha$ atoms and excluded volume were also considered. All constraints were converted into a differentiable loss function of protein atom coordinates. So the comprehensive loss could be optimized via gradient decent methods. During the optimization, the coordinates of protein atoms would fit the predicted contact map gradually. In order to simplify the description of protein structure, each residue was only represented by 6 parameters, the $C_\alpha$ coordinates and a set of Euler rotation angles (Fig. S17).

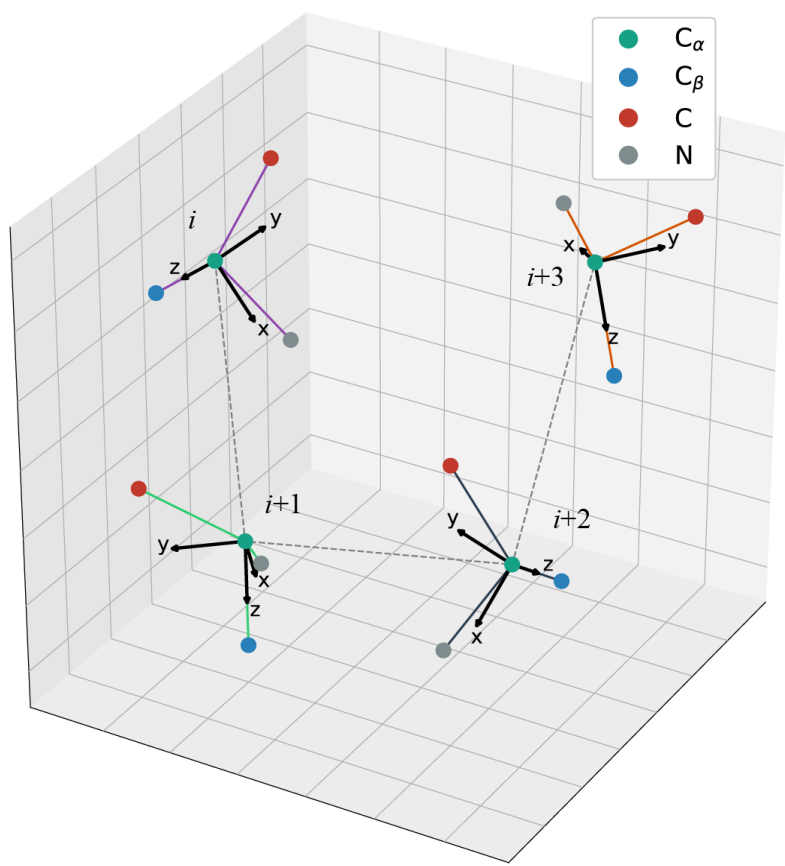

**Fig. S17. Illustration of protein structure representation.**
Each residue is located by the coordinates of the $C_\alpha$ atom, with an Euler rotation system to describe the relative positions of the other atoms. The Euler rotation system is set up by placing the $C_\alpha$ atom at the origin point, forcing the $C_\beta$ atom at the positive direction of the z-axis, and then forcing the N atom in the xOz plane.

The comprehensive loss function was minimized using a hybrid Adam-SGD optimizer. Notably, the comprehensive loss function contains many terms, each of which should be assigned a proper weight. The weights of individual loss terms were optimized on the validation set by the coordinate descent method (i.e. grid search) to avoid the bias introduced in the training set, with the performance difference of structure prediction between GDFold and CONFOLD chosen as the target optimization function.

## Availability

We prepared an online server for AmoebaContact and GDFold at the website of http://structpred.life.tsinghua.edu.cn/AmoebaContact.html.

## Funding

This work has been supported by the funds from the National Natural Science

## Author contributions

W. M. contributed to methodology, experimental design, software, formal analysis and writing of the original draft. W. D. contributed to the web server. H. G. contributed to experimental design and was responsible for supervision, writing (review and revision) as well as funding acquisition. All authors reviewed the final manuscript.

## Competing interests

The authors declare no competing interests.

# Supplementary Materials

**Details of our local contact and α-helix fragment predictor**

To provide more comprehensive contact information, we built a ResNet-based model to predict the local ($|i-j|<6$) contact information. Since the local contact prediction is much easier than the general residue contact, we just adopted the ResNet-based CNN architecture with I/R/CN (the ResNet architecture in Experiment B of Fig. S1). Based on the hyper-parameter optimization, we built a 5-unit ResNet model sharing the same input features as AmoebaContact. The number of channels and the contact cutoff were fixed at 30 and 8 Å. The corresponding F1-score on the validation set was 94.92%.

Because the protein handedness could not be distinguished from the contact/distance information. We built a multi-layer perceptron model to facilitate the structure modeling. The model will predict whether each 4-residue fragment belongs to a continuous α-helix. The 1D features (DeepCNF, SPIDER3, one-hot vectors and MSA frequency) of 4 consecutive residues and dimensionless features (protein length $L$ and the number of sequences in MSA $N$) were included as model input. The model contains 5 ResNet blocks and each layer has 30 perceptrons. The F1-score on the validation set could reach 91.4%, which is indistinguishable from other network architectures like 1D CNN.